\documentclass[longauth]{aa} 
\usepackage{graphicx}
\usepackage{longtable,lscape}

\usepackage{txfonts}
\usepackage{natbib}
\begin{document}
\title{New galactic star clusters discovered in the VVV survey. Candidates projected on the inner disk and bulge  \thanks{Based on observations gathered as part of observing programs: 179.B-2002,VIRCAM, VISTA at ESO, Paranal Observatory; NTT at ESO, La Silla Observatory (programs 087.D-0490A and 089.D-0462A) and with the SOAR telescope at the NOAO (program CN2012A-045).}}
\author{J. Borissova\inst{1,15} 
   \and
    A.-N. Chen\'{e}\inst{2}
   \and
   S. Ram\'{i}rez Alegr\'{i}a\inst{1,15}
   \and
   Saurabh Sharma\inst{3}
   \and
   J. R. A. Clarke\inst{1}
   \and
   R. Kurtev\inst{1,15}
   \and
   I. Negueruela\inst{4}
   \and
   A. Marco\inst{4}
   \and
   P. Amigo\inst{1,15}
   \and
   D. Minniti \inst{5,14,15}
   \and
   E. Bica\inst{6}
   \and
   C. Bonatto \inst{6}
   \and
   M. Catelan\inst{5,15}
   \and
   C. Fierro\inst{1}
   \and
   D. Geisler\inst{7}
   \and
   M. Gromadzki\inst{1,15}
   \and
   M. Hempel\inst{5,15}
   \and
    M. M. Hanson \inst{8}
   \and
   V. D. Ivanov\inst{9}
   \and
   P. Lucas\inst{10}
   \and
   D. Majaess\inst{11} 
   \and
   C. Moni Bidin\inst{12}  
   \and
    B. Popescu \inst{8}
   \and
   R.K. Saito\inst{13}
   }
    
\offprints{J. Borissova}

\institute{Departamento de F\'{i}sica y Astronom\'ia, Facultad de Ciencias, Universidad de Valpara\'iso, Av. Gran Breta\~na 1111, Playa Ancha, Casilla 5030, Valpara\'iso, Chile \\
\email{jura.borissova@uv.cl}
\and
Gemini Observatory, Northern Operations Centre, 670 North A’ohoku Place, Hilo, HI 96720, USA\\
\email{andrenicolas.chene@gmail.com}
\and
Aryabhatta Research Institute of Observational Sciences (ARIES), Manora Peak, Nainital, 263 129, India\\
\email{saurabh175@gmail.com}
\and
Departamento de F\'{\i}sica, Ingenier\'{\i}a de Sistemas y Teor\'{\i}a de la Se\~{n}al, Universidad de Alicante, Apdo. 99, E-03080 Alicante, Spain
\email{ignacio.negueruela@ua.es}
\and
Departamento de Astronom\'ia y Astrof\'isica, Pontificia Universidad Cat\'olica de Chile, Av. Vicu\~na Mackenna 4860, Casilla 306, Santiago 22, Chile \\ 
\email{dante@astro.puc.cl; mcatean@astro.puc.cl}
\and
Universidade Federal do Rio Grande do Sul, Departamento de Astronomia CP 15051, RS, Porto Alegre 91501-970, Brazil \\
\email{charles@if.ufrgs.br}
\and
Departamento de Astronom\'ia, Casilla 160-C, Universidad de Concepci\'on, Chile \\
\email{dgeisler@astro-udec.cl}
\and
Department of Physics,
University of Cincinnati, Cincinnati, OH 45221-0011, USA \\ 
\email{margaret.hanson@uc.edu; bogdan.popescu@uc.edu}
\and
European Southern Observatory, Ave. Alonso de Cordova 3107, Casilla 19, Santiago 19001, Chile \\
\email{vivanov@eso.org}
\and
Centre for Astrophysics Research, University of Hertfordshire, Hatfield AL10 9AB, UK \\
\email{p.w.lucas@herts.ac.uk}
\and
Department of Astronomy \& Physics, Saint Mary's University, Halifax, NS B3H 3C3;
Mount Saint Vincent University, Halifax, NS B3M 2J6, Canada\\
\email{dmajaess@ns.sympatico.ca}
\and
Instituto de Astronomia, Universidad Cat\'olica del Norte, Av. Angamos 0610, Antofagasta, Chile\\
\email{cmoni@ucn.cl}
\and
Universidade  Federal  de  Sergipe,  Departamento  de  F\'isica, Av. Marechal Rondon s/n, 49100-000, S\~ao Crist\'ov\~ao, SE, Brazil\\
\email{robsaito@gmail.com}
\and
Vatican Observatory, V00120 Vatican City State, Italy \\ 
\and
Millennium Institute of Astrophysics.
}

\date{Received; accepted}

\abstract
{VISTA Variables in the V\'{\i}a L\'actea (VVV) is one of six ESO Public Surveys using the 4 meter Visible and Infrared Survey Telescope for Astronomy (VISTA). The VVV survey  covers the Milky Way bulge and an adjacent section of the disk, and one of the principal objectives is to search for new star clusters within previously unreachable obscured parts of the Galaxy.}
{The primary motivation behind this work is to discover and analyze obscured star clusters in the direction of the inner Galactic disk and bulge.}
{Regions of the inner disk and bulge covered by the VVV survey were visually inspected using composite $JHK_{\rm S}$ color images to select new cluster candidates on the basis of apparent overdensities. DR1, DR2, CASU, and PSF photometry of $10\times10$ arcmin fields centered on each candidate cluster were used to construct color-magnitude and color-color diagrams. Follow-up spectroscopy of the brightest members of several cluster candidates was obtained in order to clarify their nature.}
{We report the discovery of 58 new infrared cluster candidates. Fundamental parameters such as age, distance, and metallicity were determined for 20 of the most populous clusters.}
{}


\authorrunning{J. \,Borissova et al.}

\titlerunning{New Galactic Star Clusters Discovered in the VVV Survey.}

\maketitle
%

\section{Introduction}

Open clusters provide important insight into the past and present properties of the Milky Way.  Open clusters feature ages that span nearly ten billion years and are detected at varying Galactocentric distances, thus enabling us to investigate the Galaxy's star formation history, the chemical evolution of the disk, and the competing influences of cluster formation and disruption that shape the properties of the current cluster population. However, to draw reliable conclusions, a sizable sample exhibiting accurately and homogeneously determined properties is desirable. The new generation of all-sky surveys (UKIDSS, the VISTA-based VHS and VVV, and Gaia) provide an opportunity to achieve those objectives. 

The Vista Variables in the V\'{\i}a L\'actea (VVV) is one of the six ESO Public Surveys using the 4 m VISTA telescope (Arnaboldi et al. 2007). The survey started in 2010, and 1929\,hours of observing time were allocated for the project over several years (Minniti et al. 2010, Saito et al. 2010, Saito et al. 2012).  The VVV survey is observing the target regions using five filters, $ZYJHK_{\rm S}$. The individual pawprints and tile images are processed by the Cambridge Astronomical Unit (CASU). Aperture photometry and astrometry are likewise performed. The VVV data are publicly available through the VISTA Science Archive (VSA, Cross et al. 2012), and additional technical information about the survey can be found in Saito et al. (2012) and Soto et al. (2013).

A primary objective of the VVV survey is to describe numerous star clusters in detail to build a statistically significant sample.  Indeed, the infrared nature of the VVV survey ensures increased sampling of clusters in highly obscured and crowded regions. The effort aims to build upon existing catalogs that are complete within only 1\,kpc from the Sun (version 3.3 - jan/10/2013 of the Dias et al. 2002 catalog; see also Lamers et al. 2005;  Piskunov et al. 2008).  In Borissova et al. (2011), we presented a catalog of 96 new cluster candidates in the disk area covered by the VVV survey. Most of these clusters were found toward known star forming regions associated with methanol maser emission, hot molecular cores (Longmore et al. 2009), Galactic bubbles detected by GLIMPSE (Churchwell et al. 2006, 2007), and infrared and radio sources.  Those tracers are indicators of early epochs of star formation. In Chen\'{e} et al. (2012) we described the methodology employed to establish cluster parameters by analyzing four known young clusters: Danks\,1, Danks\,2, RCW\,79, and DBS\,132. In Chen\'{e} et al. (2013) we presented the first study of six clusters from the Borissova et al. (2011) catalog, and at least one newly discovered Wolf-Rayet (WR) star, member of these clusters was highlighted. In Ram{\'{\i}}rez Alegr{\'{\i}}a et  al. (2014)  we presented the physical characterization of VVV CL\,086, a new massive cluster, found at the far edge of the Milky Way bar at a distance of 11$\pm6$ kpc. In this paper we report the results of our search for new star cluster candidates projected on the inner disk and bulge area covered by the VVV survey.   

\section{Cluster search and methods of analysis}

\begin{figure*}
\begin{center}
\resizebox{15cm}{!}{\includegraphics{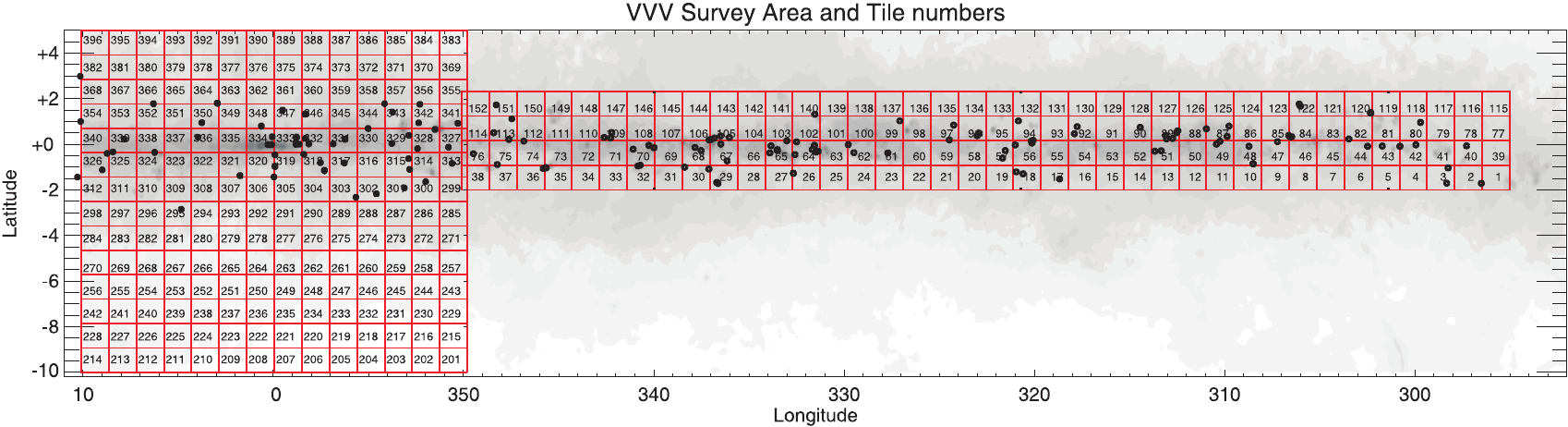}}
\caption[]{VVV survey area described by individually numbered tiles (prepared by M. Hempel,see Minniti et al. 2010 for details). The Galactic latitude {\it b} and Galactic longitude {\it l} coordinates are overlaid on a differential extinction contour map. The positions of the new star cluster candidates, 96 taken from Borissova et al. (2011) and 58 from this work are marked.
}
\end{center}
\label{vvv_field}
\end{figure*}

The VIRCAM (VISTA Infrared CAMera; Dalton et al. 2006) is a 16 detector-array infrared camera that samples 1.65\,deg$^2$. Each $2048\times2048$ detector is sensitive over $\lambda =$0.8--2.5\,$\mu$m and delivers images with an average pixel scale of 0.34\,arcsec\,px$^{-1}$.  A single exposure corresponds to a patchy individual ``pawprint'' coverage on the sky. To fill the gaps and to obtain a contiguous image, six shifted pawprints are combined into a ``tile'' covering 1.5\,deg by 1.1\,deg, which are aligned along Galactic $l$ and $b$ coordinates, respectively. To cover the VVV survey area, the disk field is subsequently divided into 152 tiles, while the bulge field contains 196 tiles (Fig.~\ref{vvv_field}). The data reduction was carried out in the canonical manner associated with infrared imaging, and details of the procedure are described by Irwin et al. (2004).

In Borissova et al. (2011) we searched for new cluster candidates by visually inspecting the true color $JHK_{\rm S}$ image tiles for obvious stellar overdensities towards known star forming regions in the Galactic disk. That procedure was repeated here, but we now include the area encompassing the Galactic bulge.  The analysis yielded 58 new cluster candidates. These are not necessarily connected with HII regions, but clearly show a significant overdensity with respect to the surrounding field. Prior to the use of color-magnitude diagrams, our main criteria for identifying star cluster candidates included a visually compact appearance that was distinct from the surrounding field, and which exhibited at least five to six stars with similar colors on the composed $J,H,K_{S}$ VVV image. Because of the subjective and qualitative method used, the catalog presented here is incomplete. 
 
 Undoubtedly, future automated searches will discover additional, less concentrated cluster candidates. However, it is likely that we uncovered the most populous objects.  In the recent compilation of stellar clusters, updated to August 2011 of all Galactic star cluster catalogs ({\it e.g.}, Dias et al. 2002; Mercer el at 2005; Dutra et al. 2003; Bica et al. 2003; Borissova et al. 2011), Morales et al. (2013) counted 2247 clusters visible in the optical and 1950 and 197 cluster candidates found in the near- and mid-infrared, respectively. Since then, Solin et al. (2012, 2014), Majaess (2013), and  Zasowski et al. (2013) reported 137; 88; 229 and 20 new embedded cluster candidates in the UKIDSS GPS, VVV, WISE, and Spitzer GLIMPSE-360 areas.  This increases the number of near-infrared cluster candidates to 2404. Froebrich et al. (2007) estimate that the spurious contamination rate (false positives) for new and unconfirmed infrared clusters from these catalogs may be as high as 50$\%$.  However, the probability of yielding {\it bona fide} stellar clusters in our sample is high (as in Borissova et al. 2011) and we expect minimal contamination from spurious detections. This is because the identified cluster candidates stem from a color-magnitude analysis (and 30\% are confirmed by spectroscopy). Our new sample is also of particular interest since it is projected against the Galactic bulge, which is a relatively challenging region. 
   
The color-magnitude diagrams (CMDs) are constructed using the First and Second Data Releases of the VISTA variables in the V\'{\i}a L\'actea (hereafter VVV DR1 and DR2; Saito et al. 2012,  \footnote{http://horus.roe.ac.uk/vsa/index.html}) and CASU\footnote{http://casu.ast.cam.ac.uk/vistasp/} catalogs, which provided aperture photometry. The tests show that the aperture radius of three pixels represents an optimum value for moderately crowded fields. For crowded regions, we performed PSF photometry of $10\times10$ arcmin fields surrounding each selected candidate. We used the VVV-SkZ pipeline, which is an automatic PSF-fitting photometric pipeline for the VVV survey (Mauro et al. 2013). Where possible, the saturated stars (usually $K_S \leq 13.5$ mag) were replaced by 2MASS stars (point source catalog).  Since 2\,MASS has a much lower angular resolution than the VVV, when replacing stars we carefully examined each cluster to avoid contamination effects of crowding. The Point Source Catalog (PSC) Quality Flags given in the 2\,MASS catalog were  used to identify reliable data. Specifically, the brightest stars are from 2\,MASS for the following clusters: VVV\,CL\,113, \,117, \,119, \,120, \,123, \,130, \,140, \,143, \,146, \,149, \,150, \,154, \,160, and CL\,161.  The typical internal photometric uncertainties of the VVV data vary from 0.009\,mag for stars with $K_{\rm S}$$\sim$13\,mag to 0.16\,mag for $K_{\rm S}$$\sim$18\,mag.

A preliminary analysis of the color-magnitude and color-color diagrams revealed 58 new star cluster candidates. Their basic properties are listed in Table~\ref{candidates}\footnote{Table~\ref{candidates} is available in electronic form (CDS), and can be accessed via anonymous ftp  (cdsarc.u-strasbg.fr , 130.79.128.5) or http://cdsweb.u-strasbg.fr/cgi-bin/qcat?J/A+A/}. The first column of the table cites the identification, followed by the equatorial coordinates of the candidate's center, the VVV tile name, a visually estimated apparent cluster radius in arcseconds, the number of probable cluster members after statistical decontamination down to $K_{S}$=16 mag, and comments on the object. Regarding the nomenclature of VVV clusters,  VVV\,CL001 is the first new VVV globular cluster candidate (Minniti et al. 2011), and the clusters VVV\,CL002, VVV\,CL003, and VVV\,CL004 were investigated by Moni Bidin et al. (2011).  The candidates from VVV\,CL005 to VVV\,CL100 were presented in Borissova et al. (2011), while the discovery and preliminary results for VVV candidates CL101, CL102, CL103, and CL105  are reported in Mauro et al. (2011), and an analysis of VVV\,CL104 is in preparation. Thus, in this paper the analysis begins with VVV\,CL106. The $JHK_{\rm S}$ composite images of all cluster candidates are shown in Appendix~1, while the individual $J$, $H$, and $K_{\rm S}$ taken from the VSA are given in Appendix~2. 

For 20 cluster candidates, we collected spectra for the brightest potential members using the IR spectrograph and imaging camera SofI in long-slit mode, which is mounted on the ESO New Technology Telescope (NTT), Chile.  The OSIRIS instrument was likewise used and is mounted on the Southern Observatory for Astrophysical Resear (SOAR) telescope, Chile. The instrument set-ups give resolution of R=2200 for SofI and 3000 \AA\ for OSIRIS. Total exposure times were typically 200--400\,s for the brightest stars and 1200\,s for the faintest. The reduction procedure for the spectra is described in Chen\'{e} et al. (2012, 2013). Spectral classification was performed using atlases of $K$-band spectra that feature spectral types stemming from optical studies (Rayner et al. 2009; Hanson et al. 1996, 2005), in concert with the spectral atlases of Martins et al. (2007), Crowther et al. (2006), Liermann et al. (2009),  Mauerhan et al. (2011), Meyer et al. (1998), and  Wallace \& Hinkle (1997). The equivalent widths (EWs) were measured from the continuum-normalized spectra using the {\sc iraf} task {\it splot}. When the S/N was high enough, the luminosity class of the star was determined using the EW of the CO line and the Davies et al. (2007) calibration. However, for spectroscopic targets displaying low S/N it was difficult to distinguish luminosity class I objects from their class III counterparts.

The procedure employed for determining the fundamental cluster parameters such as age, reddening, and distance is described in Borissova et al. (2011) and Chen\'{e} et al. (2012, 2013). As described in Borissova et al. (2011), we used the field-star decontamination algorithm of Bonatto \& Bica (2010), which was tweaked to exploit the VVV photometric depth in $H$ and $K_{\rm S}$. The first step defines a comparison field, but we need to be aware of the distribution of stars and extinction clouds in the image. To avoid biases introduced by the subjective choice of a comparison field, we selected field samples from the surrounding control fields.  The sampling geometry, position, and size of the fields were purposely varied. Since the VVV catalog is well populated, the differences in the final color-magnitude diagram from field to field were marginal. The decontamination algorithm divides the full range of magnitude and colors of a CMD into a 3D grid of cells with axes along $K_{\rm S}$, $(H-$$K_{\rm S})$, and $(J-$$K_{\rm S})$.

 Initially, cell dimensions were $\Delta$$K_{\rm S}$$=$$1.0$ and $\Delta$$(H-$$K_{\rm S})$$=$$\Delta$$(J-$$K_{\rm S})$$=$$0.2$\,mag, but sizes half and twice those values were also used. We also applied shifts in the grid positioning by $\pm1/3$ of the respective cell size along the three axes. Thus, the number of independent decontamination outputs amounted to 729 for each cluster candidate. For each cell, the algorithm estimated the expected number density of member stars by subtracting the respective field-star number density\footnote{Photometric uncertainties were taken into account by computing the probability of detecting a star of a given magnitude and color in any cell (i.e., the difference of the error function computed at the cell's borders).}. Thus, each grid setup produced a total number of member stars $N_{\rm mem}$.  Repeating the above procedure for the 729 different setups, we obtained the average number of member stars $\left<N_{\rm mem}\right>$. Each star was ranked according to the number of times it survived after all runs (survival frequency), and only the $\left<N_{\rm mem}\right>$ highest ranked stars were taken as cluster members. For the present cases we obtained survival frequencies higher than 85\%. More details about the algorithm are described in Bonatto \& Bica (2010, and references therein). Stars that are far from the main cluster fiducial lines and/or exhibit discrepant reddening and distance (spectroscopic) determinations are considered field stars and were not used to determine the mean cluster parameters. The radial velocity information, when available, was used as an additional constraint. Those results, obtained via DAOSPEC (Stetson \& Pancino 2008), warrant caution since only three to four lines were typically assessed from low-resolution and relatively noisy spectra.   

 Individual extinction and distance estimates for stars with spectral classifications were estimated using the intrinsic colors and luminosities cited by Straizys et al. (2009). The individual reddening of every star is listed in Table~\ref{param_stars}, where the uncertainties are calculated by quadratically adding the uncertainties tied to the photometry and the spectral classification (e.g., 2 subtypes). The mean values were adopted as a first guess for establishing the cluster's reddening, distance, and age via isochrone fitting. The isochrones stemmed from the Padova library (Bressan et al. 2012) and exhibited a resolution of log(Age)=0.05.
 
  For each cluster we selected the isochrones (between log(Age) from 6.6 to 10.1) corresponding to a specific metallicity, whereby the latter was derived using the Frogel et al. (2001) calibration (see below). Starting with the spectroscopically derived mean reddening, isochrones were shifted along the reddening vector from its intrinsic position in a color-color diagram (Fig.~\ref{cl111_cmd}, upper right part) until the best agreement with the observations was achieved.  The distance to the cluster was determined in the same fashion, namely by using the spectroscopic parallax as a start and shifting isochrones along the color-magnitude diagram ($K_{S}$, $J-K_{S}$).  The selection of an isochrone likewise requires knowledge of the cluster age, so the cluster distance, age, and reddening were determined iteratively.  The iterations were stopped when the parameters did not change.  Uncertainties tied to the cluster reddening and distance were calculated by accounting for the errors of the best fit, with quadratically added errors from the photometry. In certain cases (VVV CL\,119, \,143, \,149, \,150, \, 160, \,161), we used the Beletsky et al. (2009) method to estimate the age prior to isochrone fitting. The age uncertainties are conservative upper-limit estimates, which include the error of the best fit and quadratically added uncertainties of the isochrone bin.
Our cluster candidates are projected upon the bulge and inner disk, and thus we adopted the Nishiyama et al. (2009) extinction laws instead of the standard values of Cardelli et al. (1989), because it does not appear to describe observations obtained for the high-reddening regions analyzed here (see also Gonzalez et al. 2011, 2012 and Zoccali et al. 2014) as follows: $A_{J} = 1.526*E(J - K_{S})$, $A_{H} = 0.855*E(J - K_{S})$, $A_{K} = 0.528*E(J - K_{S})$,	 $A_{K} = 1.580*E(H - K_{S})$. A marginal extinction law of  R$_{V} = 2.6$ was adopted following the work of Nataf et al. (2013).

The metallicity [Fe/H] of the clusters was calculated using Eq. (3) in Frogel et al. (2001). The metallicity estimate relies on the equivalent widths of three lines: Na I (2.21 $\mu$m), Ca I (2.26 $\mu$m) atomic line blends, and the EW of the first band head of CO (2.29 $\mu$m).  Frogel et al. (2001) also calibrated the relations using more than 100 giants in 15 Galactic globular clusters with well-determined [Fe/H] values. The calibration spans a metallicity interval between -1.8 and 0 dex (on the Harris, 1996 metallicity scale). We prefer to use that calibration because it is based on moderate-resolution near-IR spectroscopy in the K band, which is similar to our spectral database. Moreover, the later work of Carrera et al. (2013) shows that the metallicity vs. spectral index relation of globular and open clusters can be successfully combined and thus the range extended between [Fe/H] +0.5 and -4.0. Consequently, we extrapolated the Frogel calibration to the metal-rich values up to +0.5 and included the uncertainties of this procedure in the error budget.  Only one star cited in Table~\ref{param_stars}  (CL\,130 Obj. 2) exhibits an abundance that is marginally beyond that range. The individual uncertainties are calculated by quadratically adding uncertainties tied to the EW measurements, the accuracy of the calibration, and the extrapolation. Generally, the metallicity cited for the clusters is calculated from an average of the most probable cluster members, and the uncertainties issued are conservative upper-limit estimates of the Poisson statistics of the measurements and quadratically added uncertainties of the individual determinations. Details of the individual cases  are given in Sec.~3, together with the number of stars used to calculate the mean. 

Recall that the radial velocity measurements for low-resolution spectra were inferred from only 3-4 lines, and should be interpreted cautiously. In general, if possible, the membership for the stars is verified on the basis of the radial velocity histogram (see the notes of individual clusters in Sec.~~3). The histogram is then fitted with a Gaussian function and the mean radial velocity emerges. The uncertainty of the mean is determined from Poisson statistics, the error of the wavelength solution and errors of the individual determination. We likewise calculated the radial velocities of field stars at the cluster's location using the Besan\c{c}on Galactic model (Robin et al. 2003).

\section{Fundamental parameters for populous cluster candidates}

In this section we present a detailed analysis of the 20 most populous star cluster candidates, by relying on photometric and spectral analyses.
  
\begin{figure}
\resizebox{9cm}{!}{\includegraphics{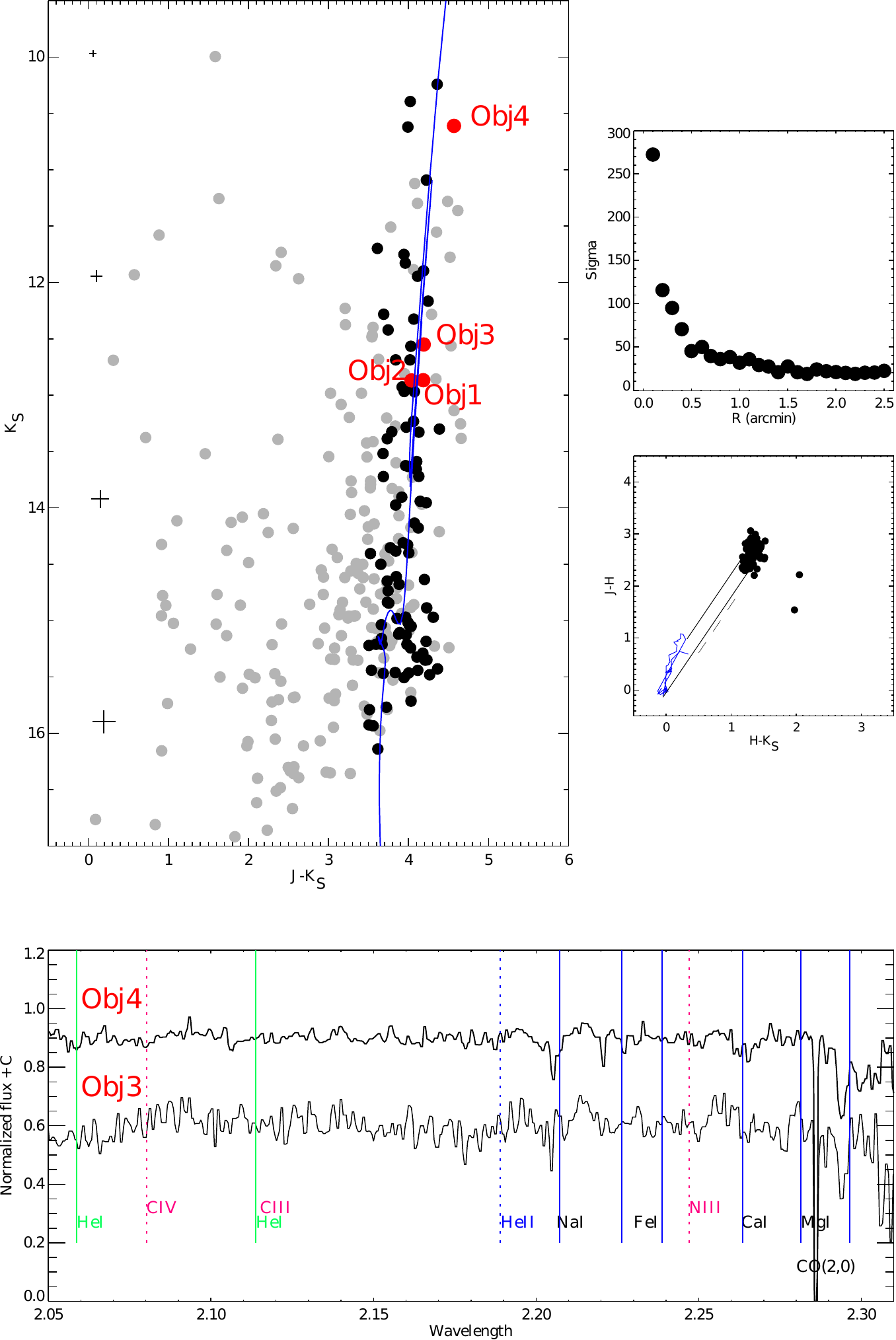}}
\caption{Top Left: $(J-K_{\rm S})$ vs. $K_{\rm S}$ color-magnitude diagram for CL\,111. Gray circles are all stars within the estimated cluster radius, dark circles are probable cluster members that remained after statistical decontamination. Stars with spectra are denoted by red circles and are labeled. The best fit is 1.6 Gyr (z=0.050) Padova isochrone (Bressan et al. 2012). The crosses convey the representative errors in a magnitude bin of 2 $K_{S}$ mag. Top right: The stellar surface density $\sigma$ (stars per square arcmin) versus radius (arcmin) of all stars in the cluster area. Bottom left: SofI low resolution spectra of Obj\,3 and Obj\,4. Bottom right: Color-color diagram of the most probable cluster members. The locus of Class III and V stars are conveyed as solid lines and are taken from Stead \& Hoare (2011, 2MASS system).}
\label{cl111_cmd}
\end{figure}

VVV\,CL111: The open cluster candidate CL\,111 lies in VVV tile b327. It is far from any known HII and bubble regions and is selected because it contains approximately 20 stars with similar colors on the composed $J,H,K_{S}$ VVV image. The cluster is relatively compact, with a radius of 35$"$. Using the CASU photometric catalog, we analyzed the $(J-K_{\rm S})$ vs. $K_{\rm S}$ diagram of the region (Fig.~\ref{cl111_cmd}). On the statistically decontaminated diagram we can identify a poorly populated red clump (RC) at $K_{\rm S}$=12.8$\pm0.1$ mag and a turn-off point (TO) at  $K_{\rm S}$=15.3$\pm0.3$ mag. The radial density profile of the cluster is shown in the right hand panel of Fig.~\ref{cl111_cmd} and the point where the radial density profile meets the level of the field stars is adopted as the projected angular size of the cluster.  The deduced result of 0.6 arcmin is consistent with that estimated visually. We observed four stars during our SofI 2012 run. Two of the spectra exhibit low S/N, and the others are classified as K0-1 and	K4-5	giants (labeled in Fig.~\ref{cl111_cmd} as Obj\,3 and Obj\,4). The analysis yields a reddening and distance modulus for the cluster of $(J-K)$=3.4$\pm0.4$ and $(M-m)_{0}$=13.1$\pm0.6$ (4.17 kpc).  The metallicity of the cluster $[Fe/H]=+0.43\pm0.2$ relies on a single star, but nonetheless agrees with the photometrically derived value of $[Fe/H]=+0.54\pm0.2$ obtained via the Ferraro et al (2004) calibration. The best Padova isochrone fit (z=0.050, Bressan et al. 2012) implies an age of 1.6$\pm0.7$ Gyr, which agrees with the result of 1.3-1.6 Gyr derived from the Beletsky et al. (2009) method. The radial velocity measurement for Obj\,4 is RV=-77$\pm$21 km/s. For comparison, the Besan\c{c}on model of the Galaxy (Robin et al. 2003) in this direction yields RV=-35$\pm$72 km/s.  Thus, this cluster candidate is most likely an old and populous metal-rich open cluster.

\begin{figure}
\resizebox{8cm}{!}{\includegraphics{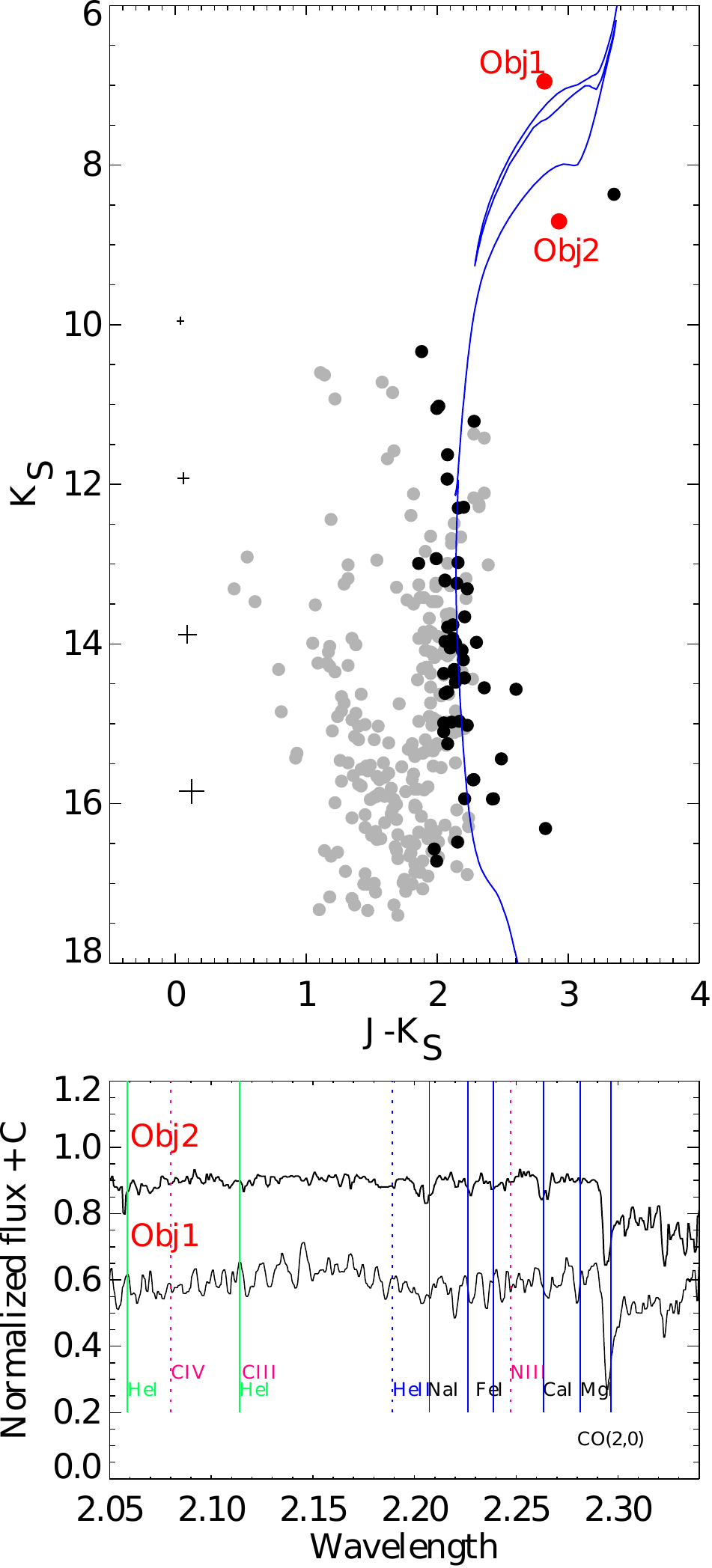}}
\caption{Top: $(J-K_{\rm S})$ vs. $K_{\rm S}$ color-magnitude diagram for CL\,113.  The symbols are the same as in Fig.~\ref{cl111_cmd}. The best fit is a 32 Myr (z=0.011) Padova isochrone, which is displayed. Bottom: SofI low resolution spectra of Obj\,1 and Obj\,2. }
\label{cl113_cmd}
\end{figure}

VVV\,CL113: This open cluster candidate lies in VVV tile b343. Approximately ten bright stars with similar colors on the composed VVV color image constitute an overdensity, within a projected radius of 20$"$. Using the VSA DR1 photometric catalog we analyzed the $(J-K_{\rm S})$ vs. $K_{\rm S}$ diagram of the region (Fig.~\ref{cl113_cmd}). Three stars are evolved from the MS, and the bulk of the probable cluster members form a poorly populated MS. Two of the brightest stars (labeled in Fig.~\ref{cl113_cmd}) were observed during our SofI 2012 run and classified as M0-1 giants.  We estimated a reddening to the cluster of $E(J-K)$=2.3$\pm0.4$, a distance modulus of $(M-m)_{0}$=13.6$\pm0.6$ (5.25 kpc), and an age of 32$\pm$7 Myr  (z=0.011). The mean metallicity of the cluster is $[Fe/H]=-0.23\pm0.15$, and the result is a mean of the velocities derived for Obj\,1 and Obj\,2 (see Table~\ref{param_stars}). Radial velocity information was only inferred from Obj\,2 (RV=+7$\pm$23 km/s), while the Besan\c{c}on model of the Galaxy cites RV=-32$\pm$96 km/s.

\begin{figure}
\resizebox{8cm}{!}{\includegraphics{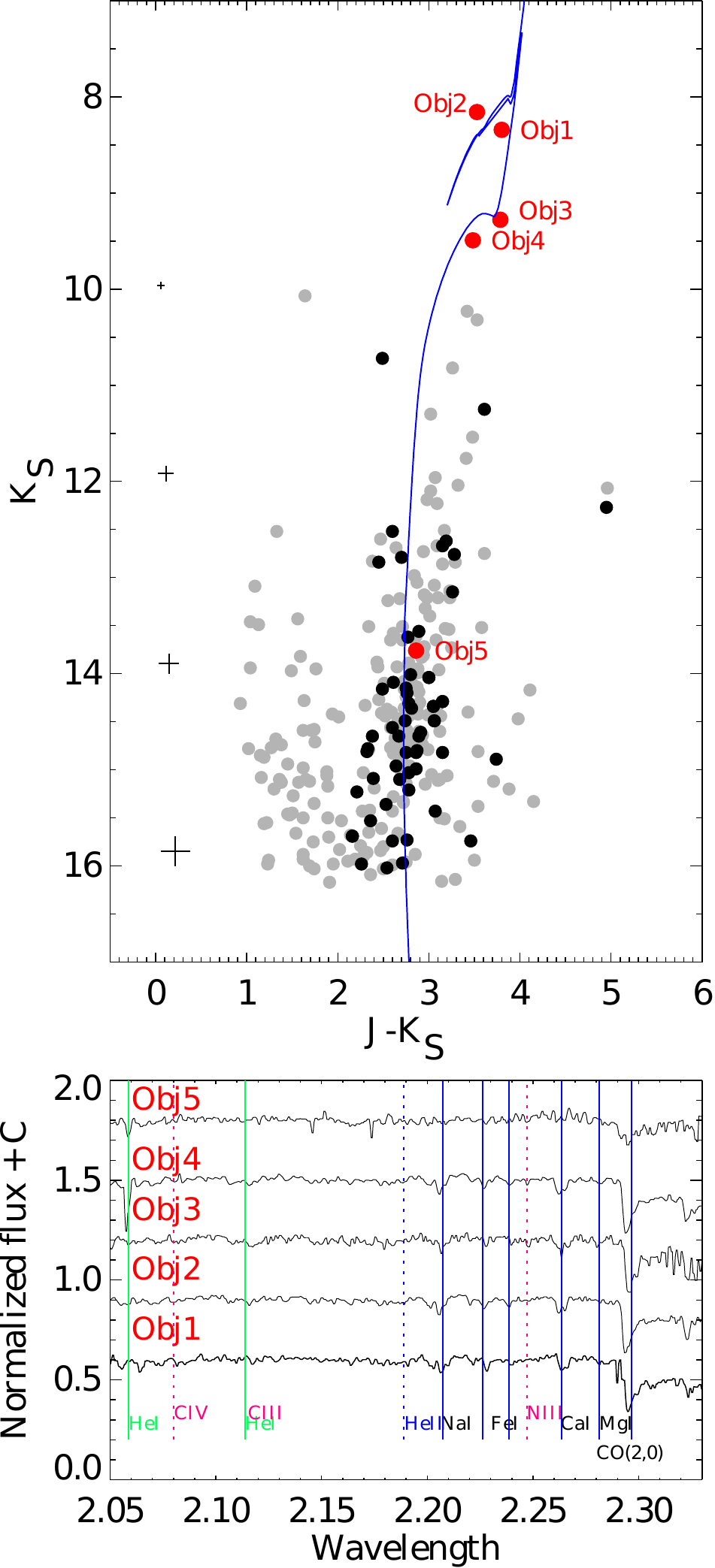}}
\caption{Top: $(J-K_{\rm S})$ vs. $K_{\rm S}$ color-magnitude diagram for CL\,117.  The symbols are the same as in Fig.~\ref{cl111_cmd}. The best fit is a 20 Myr (z=0.018) Padova isochrone, which is shown. Bottom: SofI low resolution spectra of Obj\,1, Obj\,2, Obj\,3, Obj\,4, and Obj\,5. }
\label{cl117_cmd}
\end{figure}

VVV\,CL117: This open cluster candidate lies in VVV tile b329, and an overdensity is apparent with four bright stars projected within a radius of 20$"$. These stars were observed during our SofI, 2012 run (Fig.~\ref{cl117_cmd}). Based on the spectral type vs.  EW(CO) relation established by Davies et al. (2007), the stars could be G9-K2 supergiants. The spectra of the fifth star Obj\,5 is different, however, and exhibits shallow and irregularly shaped CO and He\,I absorption lines (at $\lambda$ 2.05 $\mu$m). Interestingly,  Obj\,4 likewise displays the 2.05 $\mu$m He\,I line in absorption.  Radial velocities for the stars are highlighted in Table~\ref{param_stars}.  The RV distribution histogram fit with a Gaussian function yields a mean cluster velocity of RV=+72$\pm$25 km/s, while the Besan\c{c}on model predicts RV=-114$\pm$56 km/s. For the Obj\,5 we measured RV=+10$\pm$13 km/s, which is outside the 1 $\sigma$ interval, and is considered a field star.  We performed PSF photometry using the VVV-SkZ pipeline (Mauro et al. 2013) to construct the color-magnitude diagram. The brightest stars are clearly separated from the rest, and the color-magnitude diagram is conducive to a red supergiant cluster (RSC). The fundamental parameters are estimates as $E(J-K)$=2.9$\pm0.6$, $(M-m)_{0}$=15.35$\pm0.3$ (11.75 kpc), and age of 20$\pm$5 Myr. If confirmed, the candidate would be a rather interesting red supergiant cluster occupying a large heliocentric distance. Additional high resolution spectroscopy and deeper imaging observations are in progress and will be published in a forthcoming paper.

\begin{figure}
\resizebox{8cm}{!}{\includegraphics{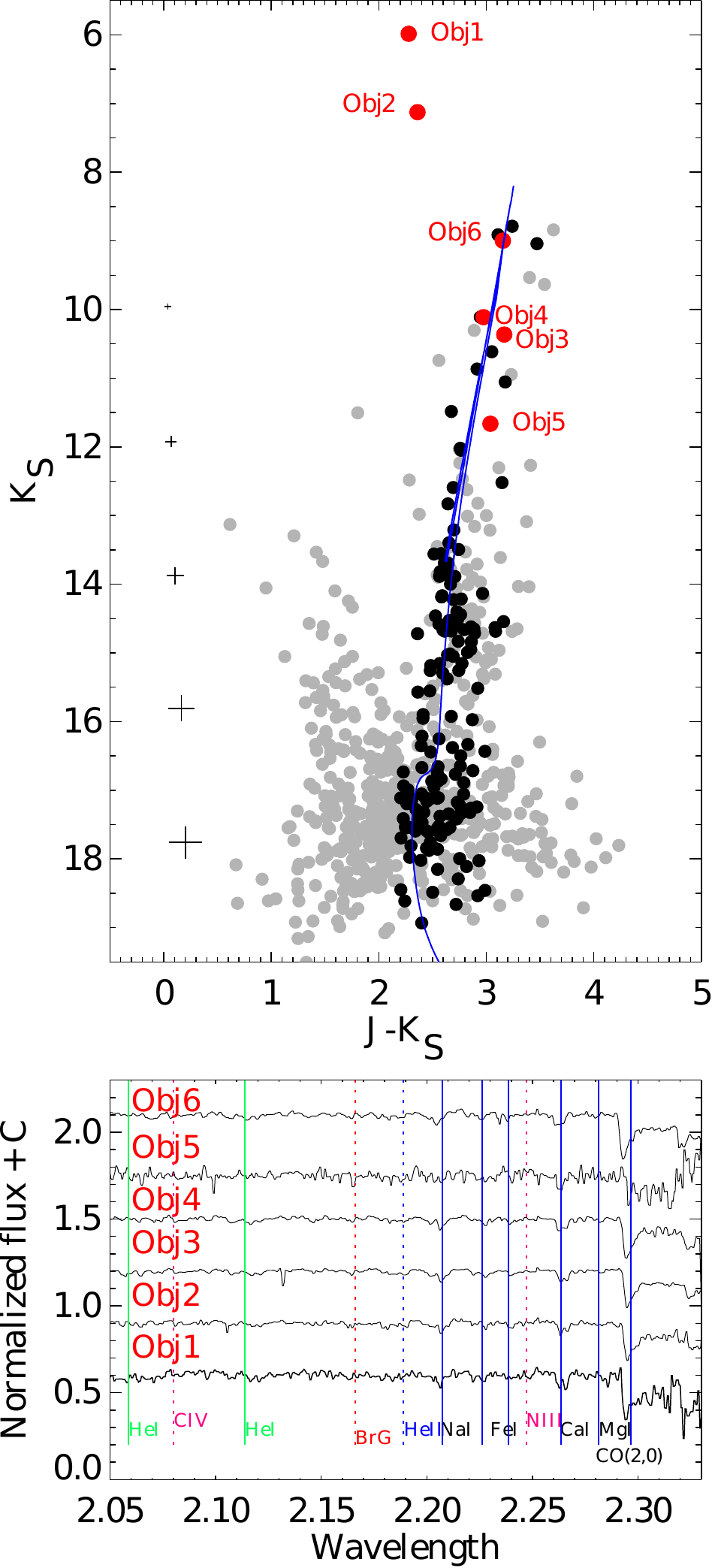}}
\caption{Top: $(J-K_{\rm S})$ vs. $K_{\rm S}$ color-magnitude diagram for CL\,119.  The symbols are the same as in Fig.~\ref{cl111_cmd}. The best fit is a 5 Gyr (z=0.009) Padova isochrone, which is displayed. Bottom: SofI low resolution spectra of Obj\,1, Obj\,2, Obj\,3, Obj\,4, Obj\,5, and Obj\,6. }
\label{cl119_cmd}
\end{figure}

VVV\,CL119: The open cluster candidate CL\,119 lies in VVV tile b344. The radius is 55$"$, which makes the cluster candidate relatively large.  Six stars were observed during our SofI, 2012 run, and are classified as K0-4 giants. The radial velocity histogram implies a mean cluster velocity of RV=+96$\pm$29 km/s, while the Besan\c{c}on model predicts RV=-77$\pm$64 km/s. All spectroscopically observed stars, except Obj\,5, lie within a 1 $\sigma$ interval, and could be considered cluster members. The brightest two stars, namely Obj\,1 and Obj\,2, are far from the mean locus of the RGB stars (Fig.~\ref{cl119_cmd}) and could be variable star candidates.   We performed PSF photometry using the VVV-SkZ pipeline (Mauro et al. 2013) to construct the color-magnitude diagram. The RC stars and the TO point are identified at $K_{\rm S}$=13.8$\pm0.2$ mag and  $K_{\rm S}$=17.3$\pm0.3$ mag, respectively. We estimate a mean reddening of $E(J-K)$=2.03$\pm0.4$, distance modulus of $(M-m)_{0}$=14.17$\pm0.3$ (6.8 kpc), and an age of 5$\pm$1.2 Gyr (z=0.009).  The mean metallicity of the cluster is calculated from an average of five stars (Obj\,5 is not included) as $[Fe/H]=-0.30\pm0.18$. Such an old cluster in the inner few kpc from the Galactic center is quite unusual, and additional observations will be acquired to investigate the candidate.

\begin{figure}
\resizebox{8cm}{!}{\includegraphics{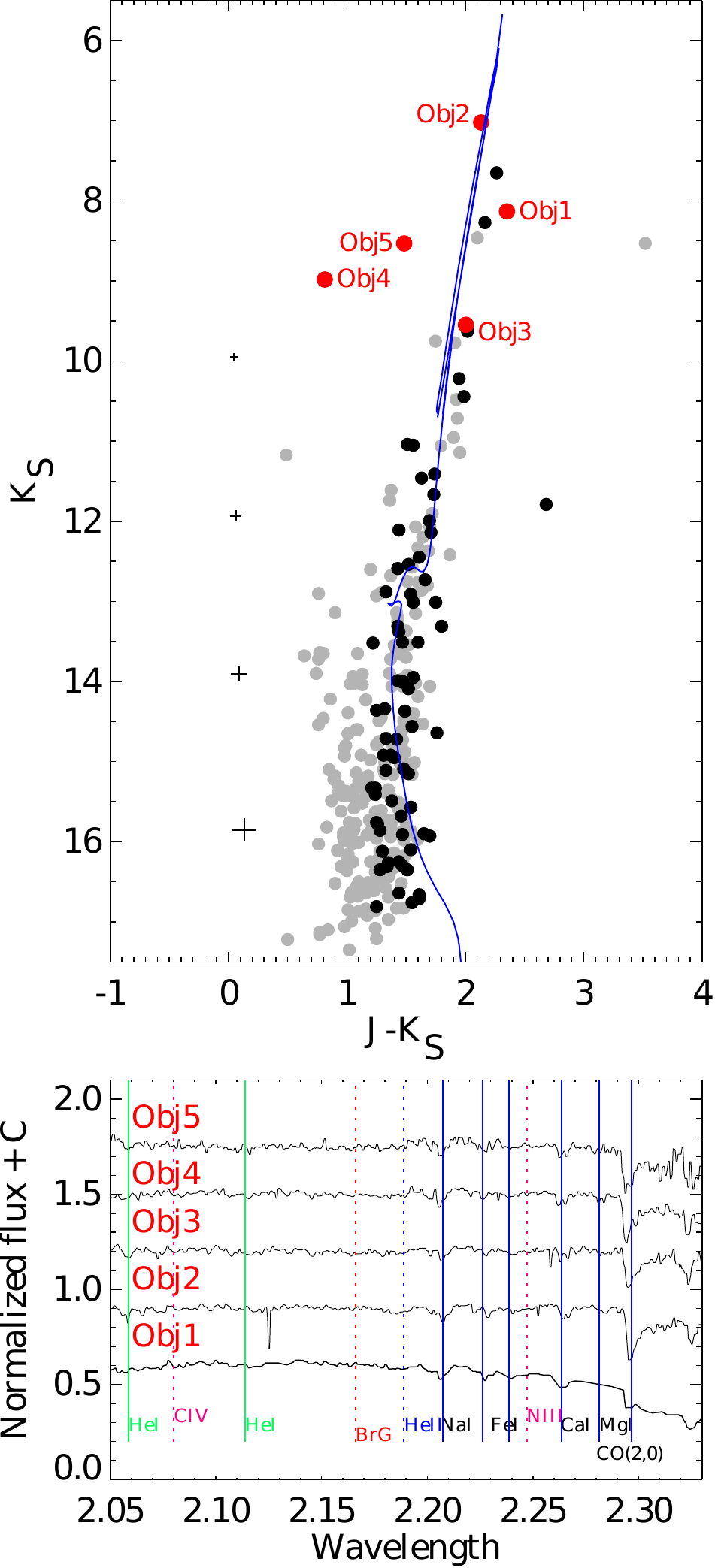}}
\caption{Top: $(J-K_{\rm S})$ vs. $K_{\rm S}$ color-magnitude diagram for CL\,120.  The symbols are the same as in  Fig.~\ref{cl111_cmd}. The best fit is a 2 Gyr (z=0.008) Padova isochrone, which is displayed. Bottom: SofI low resolution spectra of Obj\,1, Obj\,2, Obj\,3, Obj\,4, and Obj\,5. }
\label{cl120_cmd}
\end{figure}

VVV\,CL120: This open cluster candidate lies in  VVV tile b300.  The candidate constitutes a small group of relatively bright stars within a radius of 30$"$. Five stars were observed during our SofI run (Fig.~\ref{cl120_cmd}) and were classified as K5-M2 giants. The VVV DR1 database was used to construct the color-magnitude diagram. The RGB sequence is visible, and main sequence (MS) stars can likewise be seen.  Objects 4 and 5 appear distant from the mean locus of the RGB stars, and exhibit lower reddening and different radial velocities (see Table~\ref{param_stars}).  Those objects are probably field stars. The radial velocity of the cluster is RV=+51$\pm$9 km/s and stems from a mean of Obj\,2 and 3, while the Besan\c{c}on model predicts RV=-16$\pm$59 km/s in this direction. The mean reddening and distance modulus are $E(J-K)=1.2\pm0.1$ and $(M-m)_{0}$=11.6$\pm0.6$ (2.09 kpc), respectively. The best fit Padova isochrone (z=0.008) implies a cluster that is 2$\pm$0.5 Gyr old. The mean metallicity of the cluster is $[Fe/H]=-0.36\pm0.45$, and is merely the average of the metallicites of Obj\,2 and Obj\,3.

\begin{figure}
\resizebox{8cm}{!}{\includegraphics{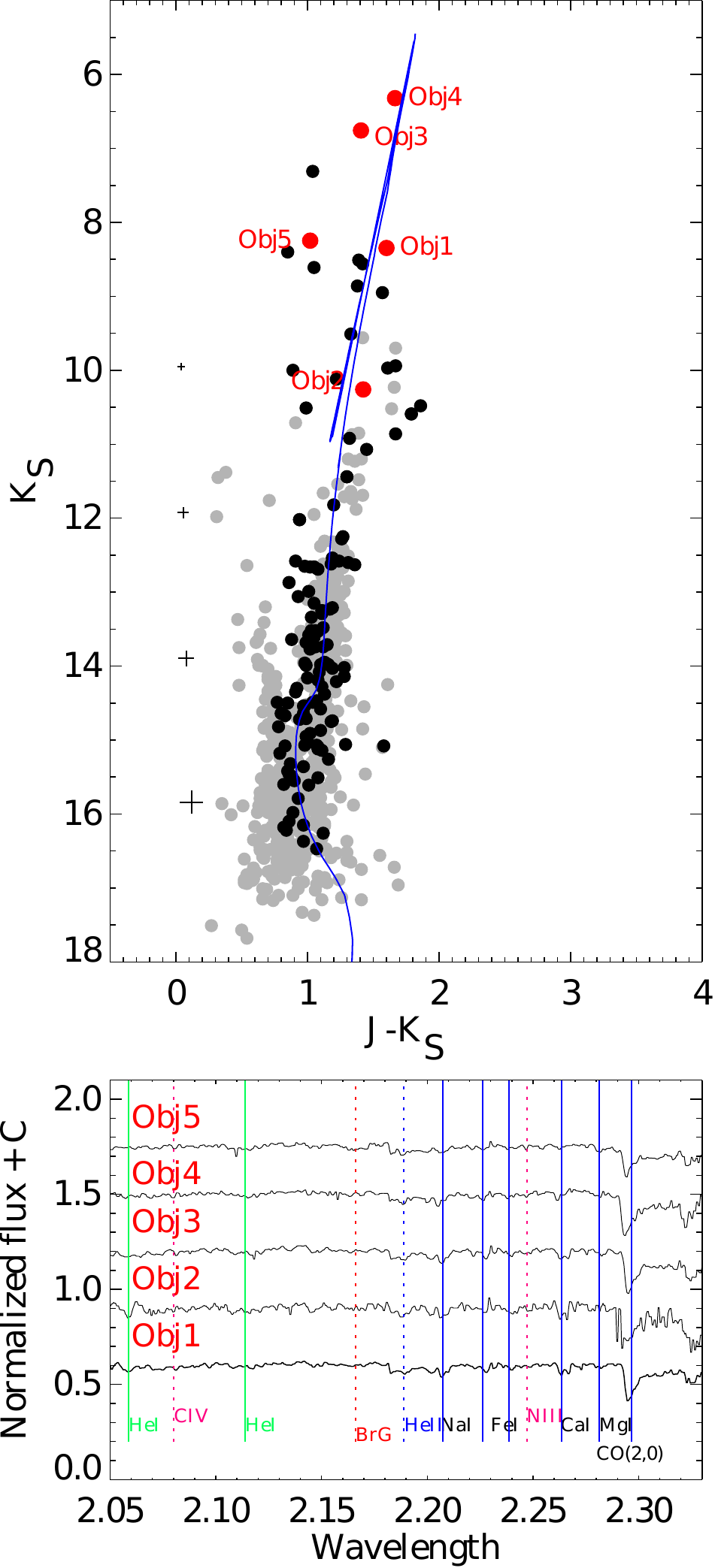}}
\caption{Top: $(J-K_{\rm S})$ vs. $K_{\rm S}$ color-magnitude diagram for CL\,123.  The symbols are the same as in  Fig.~\ref{cl111_cmd}. The best fit is a 9 Gyr (z=0.012) Padova isochrone, which is shown. Bottom: SofI low resolution spectra of Obj\,1, Obj\,2, Obj\,3, Obj\,4 and, Obj\,5. }
\label{cl123_cmd}
\end{figure}

VVV\,CL123: This open cluster candidate lies in VVV tile b301. Several bright stars form an overdensity within a 55$"$ radius. Five stars were observed in our SofI 2012 run (Fig.~\ref{cl123_cmd}). They are classified as G9-M6 giants. The VVV DR2 was used to construct the color-magnitude diagram. The RGB sequence is visible, in concert with main sequence stars.  The TO point lies near $K_{s}$=15.0$\pm0.3$.  Object 5 is situated far from the mean locus of the RGB stars and features a different velocity which implies it is likely a field star (see Table~\ref{param_stars}). The S/N of Obj\,2 is low, and consequently a radial velocity measurement could not be secured. The mean radial velocity is RV=-50$\pm$17 km/s, based on Obj\,1, Obj\,3, and Obj\,4.  The Besan\c{c}on model predicts RV=-20$\pm$78 km/s along this direction. We estimated a reddening of $E(J-K)=0.55\pm0.21$ and distance modulus of $(M-m)_{0}$=12.1$\pm0.5$ (2.63 kpc). The best Padova isochrone fit yields a cluster age of 9$\pm$0.7 Gyr (z=0.012). The mean metallicity of the cluster is $[Fe/H]=-0.19\pm0.14$, and relies on the the individual metallicities of Obj\,1, Obj\,2, Obj\,3, and Obj\,4.

\begin{figure}
\resizebox{8cm}{!}{\includegraphics{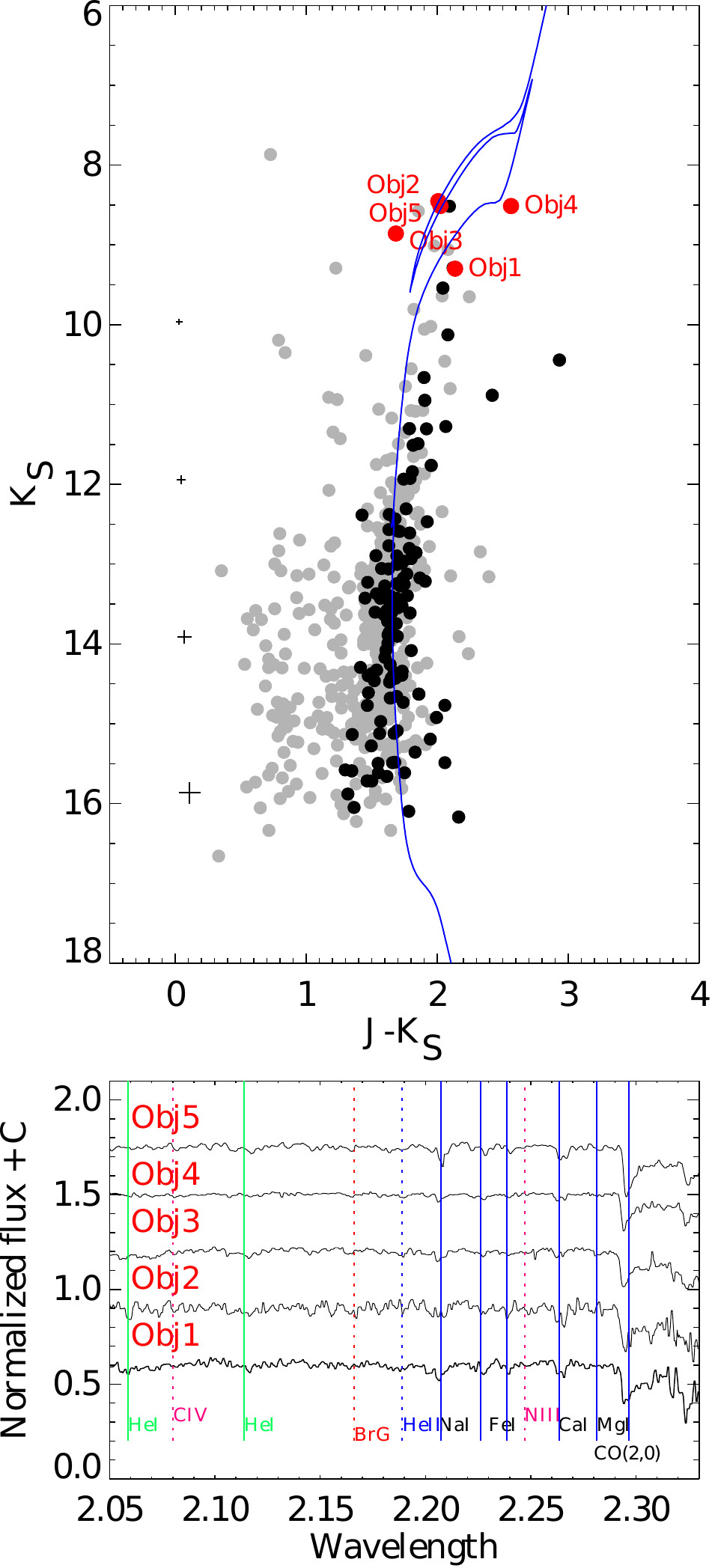}}
\caption{Top: $(J-K_{\rm S})$ vs. $K_{\rm S}$ color-magnitude diagram for CL\,124.  The symbols are the same as in  Fig.~\ref{cl111_cmd}. The best fit is a 50 Myr (z=0.007) Padova isochrone, which is shown. Bottom: SofI low resolution spectra of Obj\,1, Obj\,2, Obj\,3, Obj\,4, and Obj\,5. }
\label{cl124_cmd}
\end{figure}

VVV\,CL124: This open cluster candidate lies in the VVV tile b346. The candidate appears as a concentrated group of seven to eight stars within a radius of 15$"$. Five stars were observed with SofI (Fig.~\ref{cl124_cmd}) and are classified as K1-M1 giants. The VVV DR2 database was used to construct the color-magnitude diagram, which shows a distinct sequence of evolved and main sequence stars.  We estimated the reddening to be $E(J-K)$=1.8$\pm0.3$ and the distance is $(M-m)_{0}$=13.6$\pm0.5$ (5.25 kpc). The best Padova isochrone fit implies an age of  50$\pm$6 Myr (z=0.007). The radial velocity distribution displays comparable velocities for all five stars (see Table~\ref{param_stars}), with a mean value of RV=-36$\pm$8 km/s.  The Besan\c{c}on model predicts RV=-18$\pm$103 km/s in this direction. The mean metallicity is $[Fe/H]=-0.4\pm0.5$, and was determined using all the stars. 

\begin{figure}
\resizebox{8cm}{!}{\includegraphics{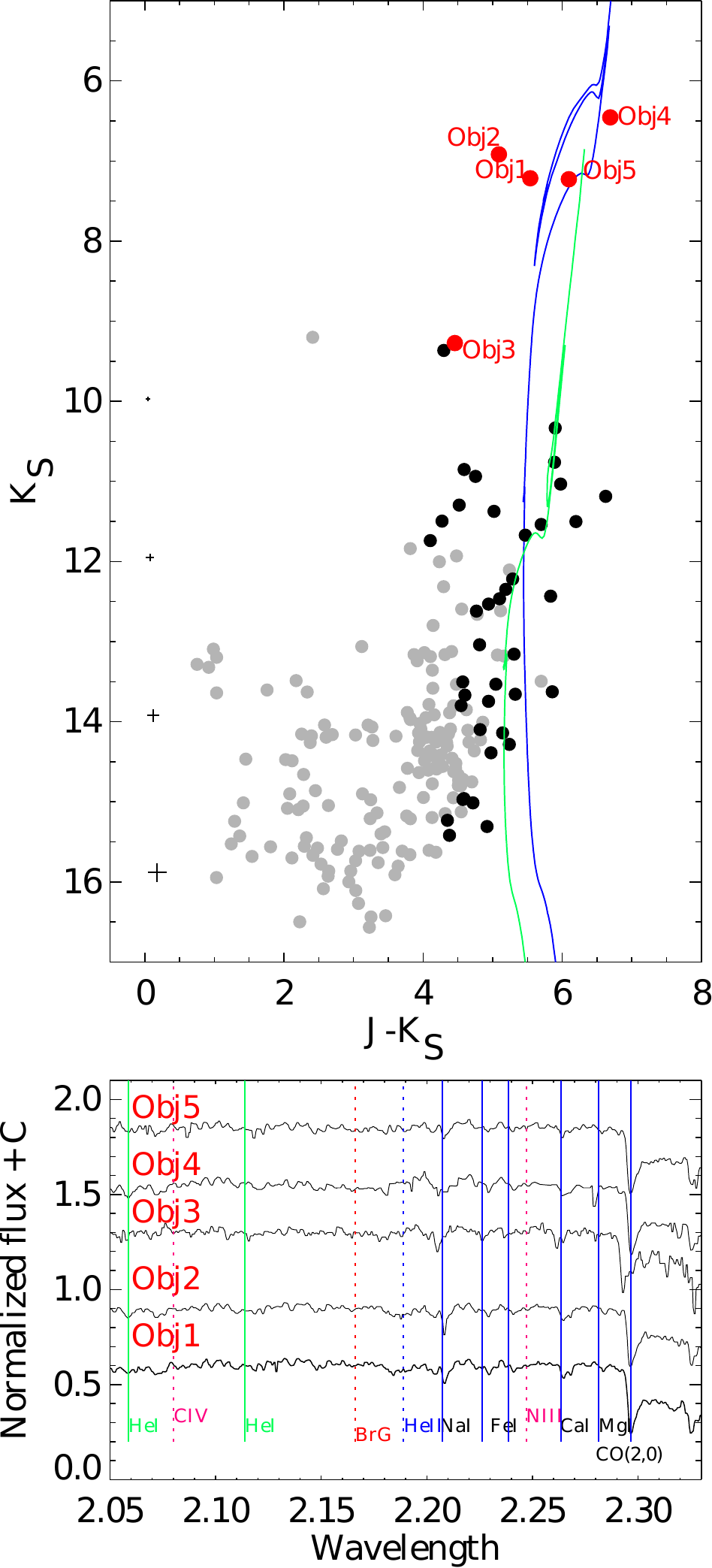}}
\caption{Top: $(J-K_{\rm S})$ vs. $K_{\rm S}$ color-magnitude diagram for CL\,130.  The symbols are the same as in  Fig.~\ref{cl111_cmd}. The best fits are 32 (blue) and 316 Myr (green) Padova isochrones (z=0.018), which are overplotted. Bottom: SofI low resolution spectra of Obj\,1, Obj\,2, Obj\,3, Obj\,4, and Obj\,5. }
\label{cl130_cmd}
\end{figure}

VVV\,CL130: This open cluster candidate lies in VVV tile b332. Similar to the case of VVV CL\,117, four bright stars are concentrated within a radius of  20$"$. These stars were observed with SofI during our 2012 run (Fig.~\ref{cl130_cmd}). Based on the spectral type vs. EW(CO) relation established by Davies et al. (2007), the stars could be classified as G9-K2 supergiants. The position in the CMD of  Obj\,3 suggests that it is probably a field star. The radial velocity histogram exhibits a peak near RV=-20$\pm$8 km/s, while the Besan\c{c}on model predicts RV=-12$\pm$76 km/s. We performed PSF photometry using the VVV-SkZ pipeline (Mauro et al. 2013) to construct a color-magnitude diagram. The brightest stars are separated from the rest in the upper part of the color-magnitude diagram, which is typical for RSG clusters. There is no indication of main sequence stars in our photometry. Assuming that the candidate is an RSG cluster, we estimated a reddening of $E(J-K)=5.6\pm0.3$, distance modulus of $(M-m)_{0}$=11.0$\pm0.4$ (1.58 kpc), and age of 32$\pm$8 Myr (z=0.018). If confirmed, it would be an interesting and heavily reddened RSG cluster projected close to the Sun.  However, we cannot exclude a luminosity class III classification outright , and in that instance the best fit yields $E(J-K)=5.2\pm0.3$, $(M-m)_{0}$=11.4$\pm0.4$ (1.9 kpc), and age of 316 $\pm$37 Myr.  Additional observations are being obtained and will be published in a subsequent paper. 

\begin{figure}
\resizebox{9cm}{!}{\includegraphics{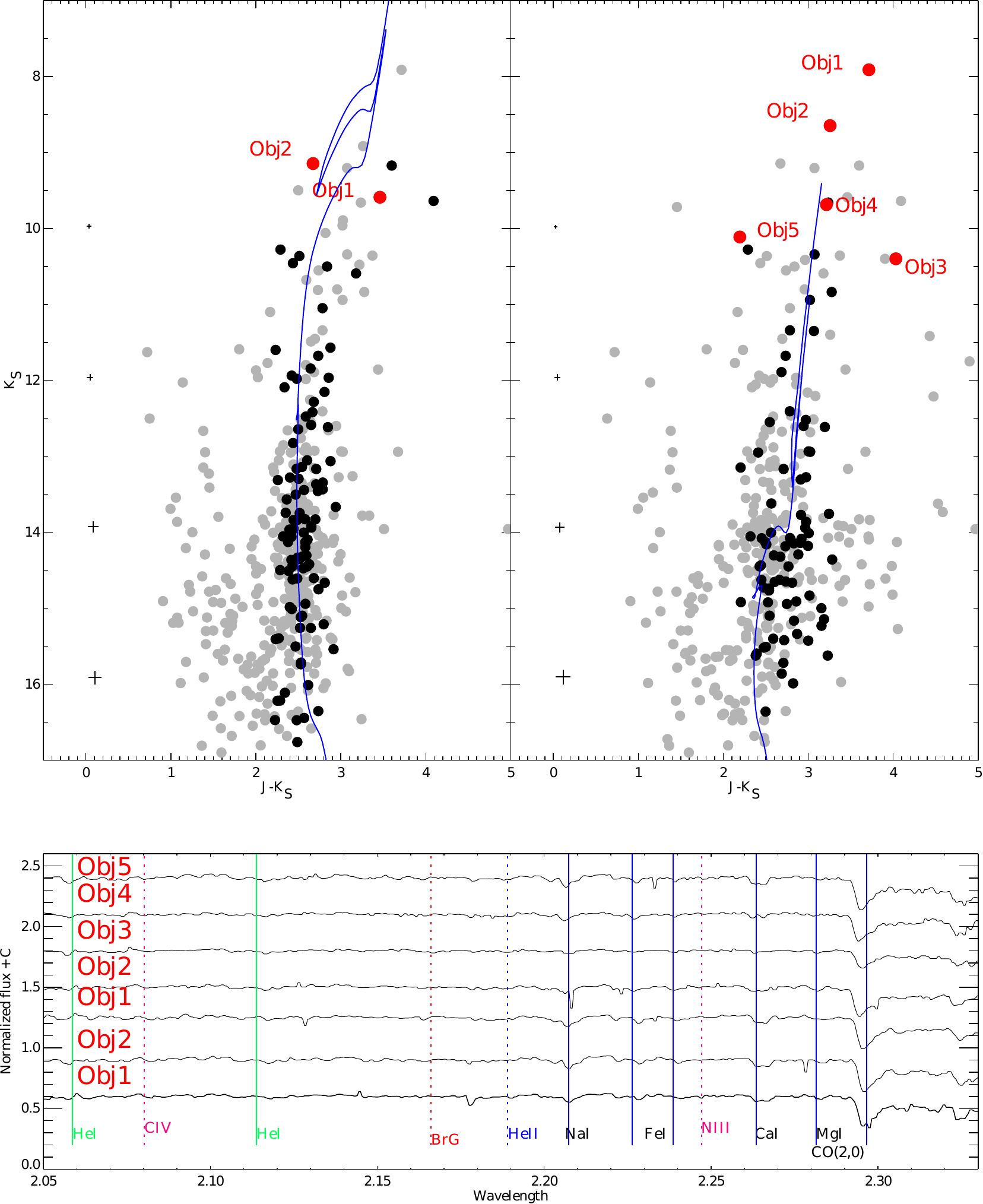}}
\caption{Top: $(J-K_{\rm S})$ vs. $K_{\rm S}$ color-magnitude diagram for CL\,139 (left) and CL\,140 (right).  The symbols are the same as in  Fig.~\ref{cl111_cmd}. The best fits are 80 Myr and 1.3 Gyr Padova isochrones (Bressan et al. 2012), respectively. Those isochrones are tied to z=0.012 and z=0.008 abundances, accordingly.  Bottom: SofI low resolution spectra of Obj\,1 and Obj\,2 of CL\,139 and Obj\,1, Obj\,2, Obj\,3, Obj\,4 and Obj\,5, of CL\,140. }
\label{cl139_cmd}
\end{figure}

VVV\,CL139 and VVV\,CL140: These two cluster candidates lie in VVV tile b318. The two groups are projected close together (less than 1 arcmin) and form an overdensity with respect to the field. Interactions between star clusters are short processes, which may lead to star cluster disruption (i.e., infant mortality) or to a merger scenario.  If the pair is  gravitationally bound, they may form a more massive cluster, mixing the constituent stellar generations.  CL\,139 and CL\,140 could be two independent clusters simply constituting a spatial projection along the sight line or potentially a single cluster divided owing to dust extinction. To verify the last hypothesis we retrieved WISE, GLIMPSE and MIPSGAL 24 $\mu$m images. No signatures of strong dust emission between the cluster candidates can be found, and the images show a relatively homogeneous background.  Marginal dust is visible to the left of CL\,140, which could partially obscure the cluster. Two stars in CL\,139 and five stars in CL\,140 have been observed with SofI during our 2012 observing run. The stars are plotted in Fig.~\ref{cl139_cmd}.  All are late type G9-M1 giants. Radial velocities cited in Table~\ref{param_stars} are comparable within the uncertainties. The VVV DR1 database was used to construct the color-magnitude diagrams (Fig.~\ref{cl139_cmd}).  The cluster candidate CL\,139 displays evolved stars and a distinct MS. We estimate cluster parameters of $E(J-K)$=$2.6\pm0.3$ and $(M-m)_{0}$=12.90$\pm0.6$ (3.8 kpc). The age of this cluster candidate is approximately 80$\pm$19 Myr. The mean metallicity of the cluster is $[Fe/H]=-0.20\pm0.25$ and is linked to Obj\,1 and Obj\,2. The cluster candidate CL\,140 exhibits a distinct RGB, and the TO point discernible at $K_{s}$=14.0$\pm0.1$. The cluster is slightly less reddened ($E(J-K)$=$2.3\pm0.4$) than CL\,139. An isochrone of 1.3$\pm$0.3 Gyr fits the CMD, assuming the above distance. The results are consistent with the mean metallicity of the cluster estimated as  $[Fe/H]=-0.42\pm0.35$ (Obj\,1, 2, and 4).  The existing data imply that a new possible cluster pair in the Galaxy has been discovered.
 
\begin{figure}
\resizebox{8cm}{!}{\includegraphics{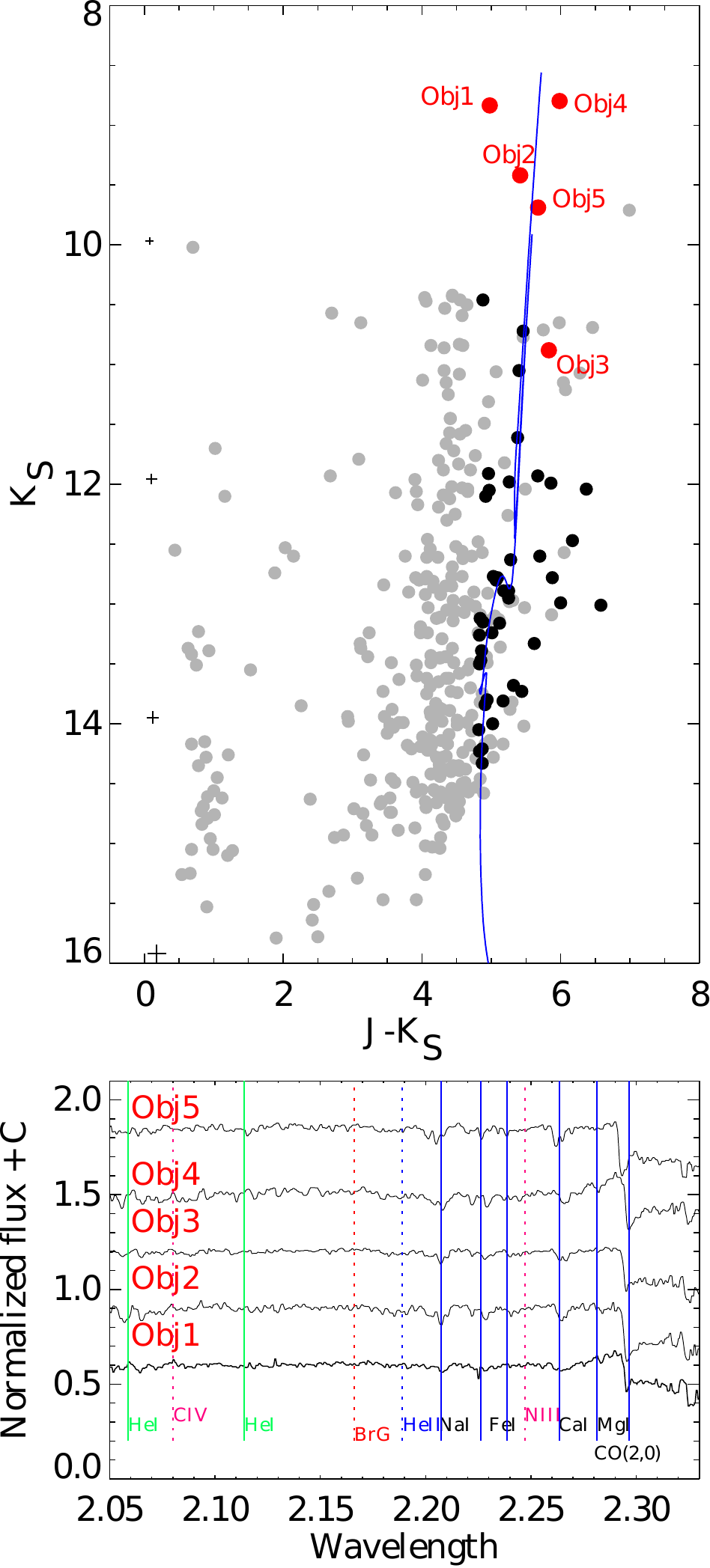}}
\caption{Top: $(J-K_{\rm S})$ vs. $K_{\rm S}$ color-magnitude diagram for CL\,142.  The symbols are the same as in Fig.~\ref{cl111_cmd}. The best fit is an 800 Myr (z=0.013) Padova isochrone, which is plotted. Bottom: SofI low resolution spectra of Obj\,1, Obj\,2, Obj\,3, Obj\,4, and Obj\,5.}
\label{cl142_cmd}
\end{figure}

VVV\,CL142: The VVV\,CL142 star cluster candidate lies in VVV tile b333, and occupies a complex field encompassed by dark clouds according to the MIPSGAL 24 $\mu$m image. The cluster candidate appears as a compact group (radius of 20$"$, Fig.~\ref{cl142_cmd}) of 6 stars with similar colors, as inferred from the VVV images. Five of those targets were observed and subsequently classified as G3-M5 giants. The radial velocity of Obj\,1 is larger than that of Obj\,4 and Obj\,5, which in concert with its CMD position, indicates this star could belong to the field. The mean velocity of Obj\,4 and Obj\,5 is RV=14$\pm17$ km/s, while the Besan\c{c}on model predicts -11$\pm70$ km/s.  A similar value is reported in the velocity map of the Galactic bulge giants by Zoccali et al. (2014).  The VVV DR1 photometry reveal RGB stars and a poorly populated MS. The estimated cluster parameters are $E(J-K)$=$4.8\pm0.3$ and $(M-m)_{0}$=$11.25\pm0.6$ (1.8 kpc).  The cluster age of  800$\pm$92 Myr (z=0.013) was determined using a Padova isochrone (Bressan et al. 2012). The mean metallicity  ($[Fe/H]=-0.14\pm0.4$) was computed using all the spectroscopically observed stars. If confirmed, this would be another case of a highly reddened intermediate age cluster discovered relatively close to the Sun.

\begin{figure}
\resizebox{8cm}{!}{\includegraphics{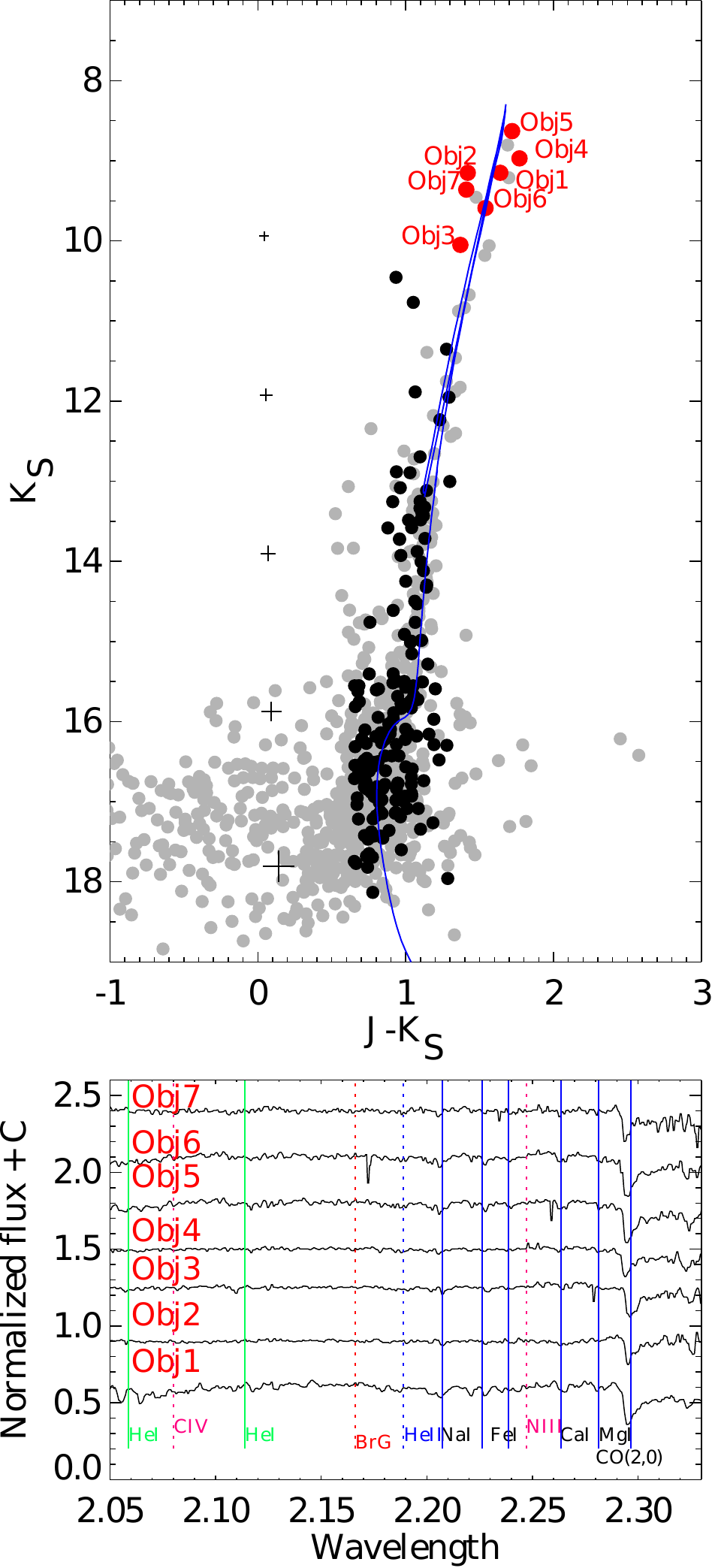}}
\caption{Top: $(J-K_{\rm S})$ vs. $K_{\rm S}$ color-magnitude diagram for CL\,143.  The symbols are the same as in  Fig.~\ref{cl111_cmd}, and probable candidates within 25 arcsec are shown. The best fit is a 4 Gyr (z=0.004) Padova isochrone, which is displayed. Bottom: SofI low resolution spectra of Obj\,1, Obj\,2, Obj\,3, Obj\,4, Obj\,5, Obj\,6, and  Obj\,7. }
\label{cl143_cmd}
\end{figure}

VVV CL\,143: The cluster candidate VVV CL\,143 lies in VVV tile b302, and appears as a group of 10-12 relatively bright stars within a radius of 35$"$.  Seven stars were observed during our SofI run (Fig.~\ref{cl143_cmd}), and exhibit similar velocities (see Table~\ref{param_stars}) with a mean of RV=+86$\pm26$ km/s.  Conversely, the Besan\c{c}on model predicts -126$\pm67$ km/s.  The stars are classified as G8-M1 giants. We performed PSF photometry using VVV-SkZ pipeline (Mauro et al. 2013) to construct the color-magnitude diagram. The diagram conveys a defined RGB, several RC stars at $K_{\rm S}$=13.3$\pm0.2$ mag, and the TO at $K_{\rm S}$=16.62$\pm0.3$ mag.  We estimated cluster parameters of $E(J-K)$=0.58$\pm0.2$, $(M-m)_{0}$=14.45$\pm0.6$ (7.8 kpc), and age of 4$\pm$0.7 Gyr.  The mean metallicity of the cluster is $[Fe/H]=-0.62\pm0.52$, which was evaluated using all spectroscopically observed stars. The data could imply that the candidate is a populated old open cluster. However, taking the morphology of the CMD and its relatively low metallicity into account, it could be a young globular cluster. Additional support for that hypothesis is the large number of bright objects near the top of the isochrone of CL\,143. Those objects could be AGB stars since the red clump appears well populated and the turn-off is defined.  The position of the cluster in the Galaxy should also be considered, and old open clusters are rarely detected in the inner Galaxy. Friel (1995) argued that they should be completely absent in the inner 7 kpc and, yet a globular cluster in the location of CL\,143 would be among the youngest that can have an extragalactic origin (Moni Bidin et al. 2011).  A detailed analysis is necessary to clarify the nature of this interesting cluster candidate. 
 
\begin{figure}
\resizebox{8cm}{!}{\includegraphics{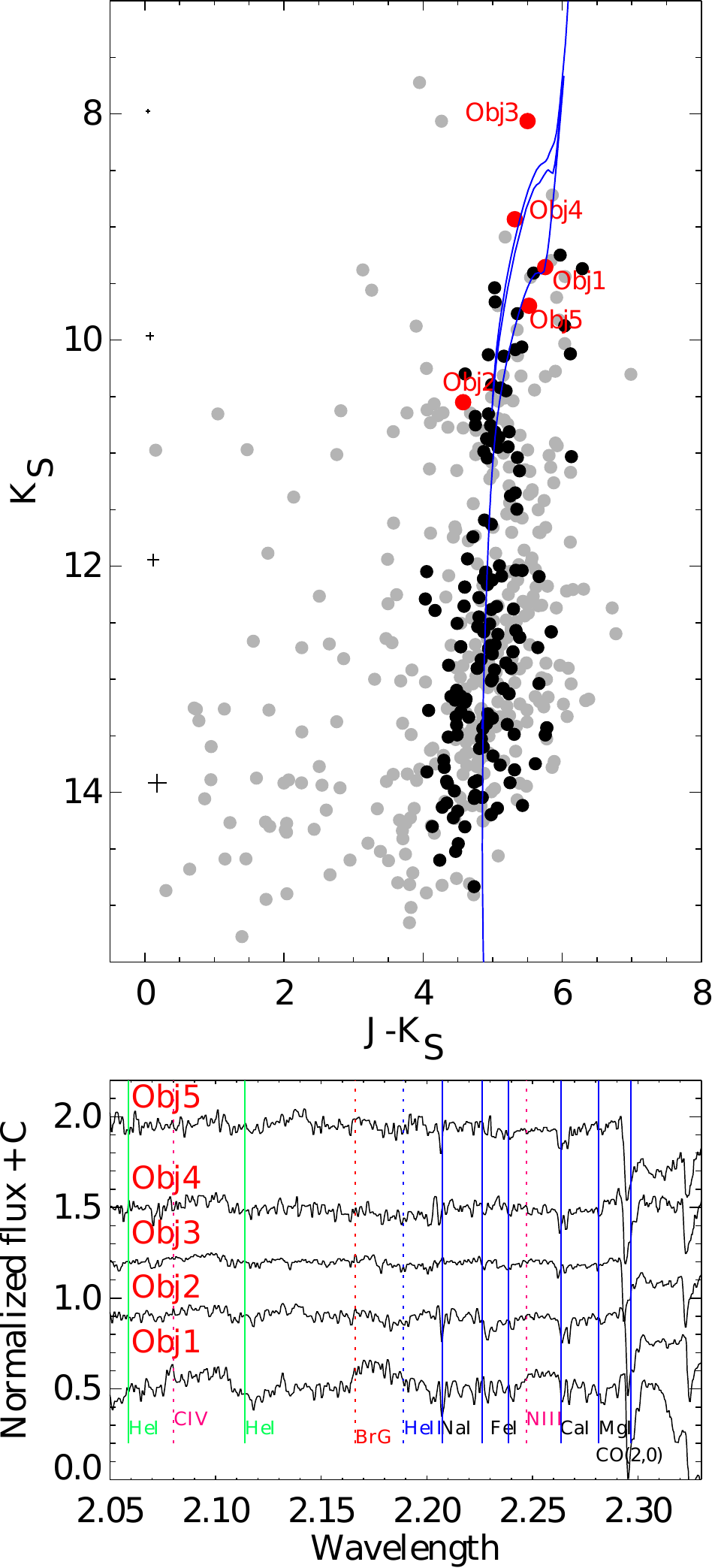}}
\caption{Top: $(J-K_{\rm S})$ vs. $K_{\rm S}$ color-magnitude diagram for CL\,146.  The symbols are the same as in Fig.~\ref{cl111_cmd}. The best fit is a 40 Myr (z=0.013) Padova isochrone, which is shown. Bottom: SofI low resolution spectra of Obj\,1, Obj\,2, Obj\,3, Obj\,4, and Obj\,5. }
\label{cl146_cmd}
\end{figure}

VVV CL\,146: The cluster candidate VVV CL\,146 lies in tile b333, very close to the galactic center, and appears as a loose group of stars within a radius of 30$"$. Five stars were observed with SOAR, 2012 run (Fig.~\ref{cl146_cmd}) and were classified as K5-M7 giants.  The VVV DR1 color-magnitude diagram reveals some evolved RGB and MS stars. It is extremely hard to analyze this region of the galaxy because of heavy crowding and differential reddening. The WISE and MIPSGAL images show obscuring clouds of dust. We have estimated the reddening of $E(J-K_{\rm S})$=$5.0\pm0.5$ and distance modulus of $(M-m)_{0}$=$13.2\pm0.6$ (4.37 kpc). 
 The radial velocity histogram exhibits a peak near RV=123$\pm$24 km/s, while the Besan\c{c}on model predicts RV=-11$\pm$75 km/s.
 The age of this star cluster candidate is estimated to be around 40$\pm$7 Myr for z=0.013 Padova isochrone. The mean metallicity of the cluster is calculated as  an average of all spectroscopically observed stars as $[Fe/H]=-0.15\pm0.25$.  Deeper photometry is needed to confirm the reality of this cluster candidate.

\begin{figure}
\resizebox{8cm}{!}{\includegraphics{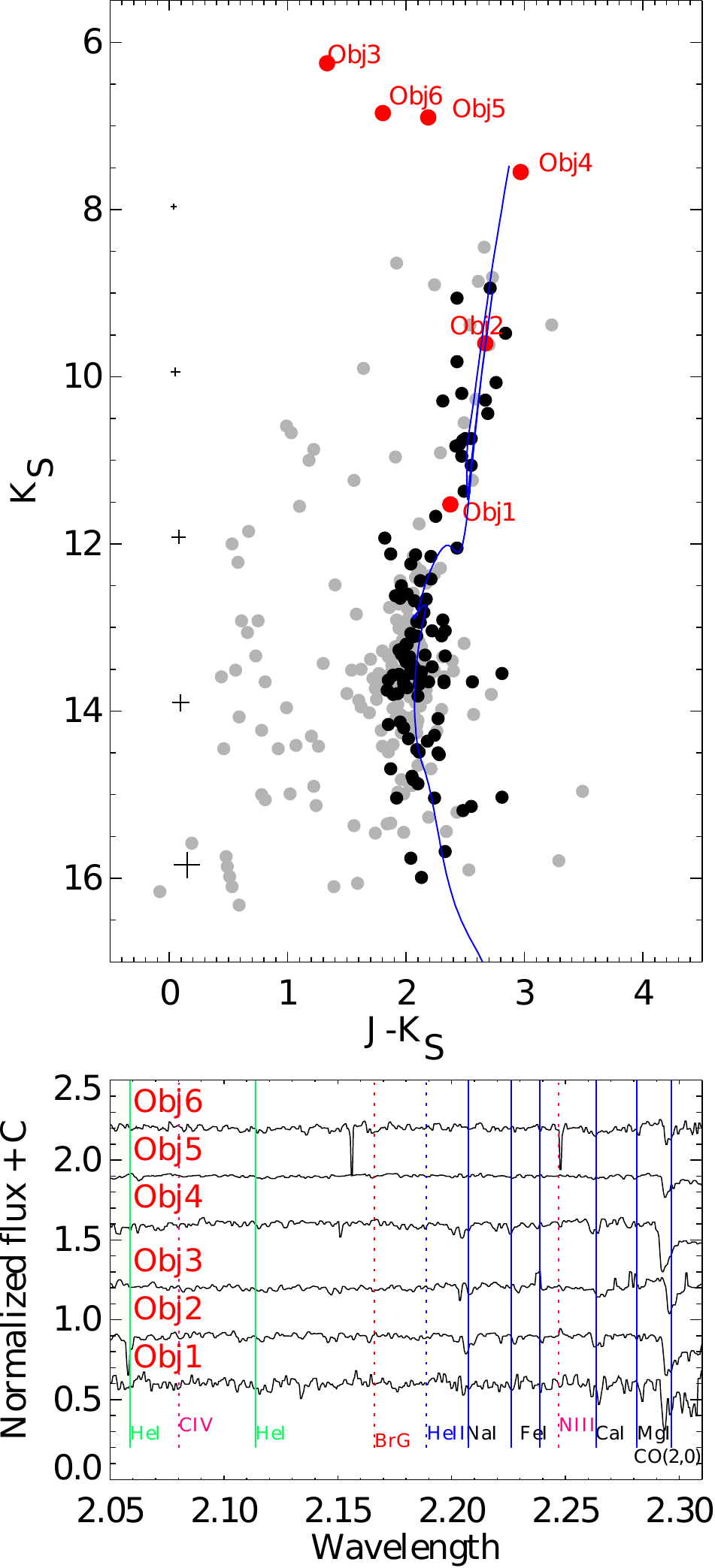}}
\caption{Top: $(J-K_{\rm S})$ vs. $K_{\rm S}$ color-magnitude diagram for CL\,149.  The symbols are the same as in Fig.~\ref{cl111_cmd}. The best fit is a 1.3 Gyr (z=0.007) Padova isochrone, which is shown. Bottom: SofI low resolution spectra of Obj\,1, Obj\,2, Obj\,3, Obj\,4, Obj\,5, and Obj\,6.}
\label{cl149_cmd}
\end{figure}

VVV CL\,149: The cluster candidate VVV CL\,149 lies in VVV tile b319. The surrounding area contains numerous infrared sources discovered by Ojha el al. (2003) using ISOGAL. Several stars form a compact group projected within a 30$"$ radius. Consequently, we initially suspected that this was a young star cluster candidate. Using SofI and the NTT-ESO telescope we observed six stars during our 2012 run (Fig.~\ref{cl149_cmd}).  All stars exhibit CO lines and are classified as G9-M0 giants.  The VVV DR1 color-magnitude diagram reveals RGB and main sequence stars. The TO point appears near $K_{\rm S}$=12.6$\pm0.2$ mag. The brightest stars, Obj\,3, Obj\,5, and Obj\,6, are far from the mean locus of the RGB population.  The targets are likely field stars, which is likewise supported by their RV velocities (see Table~\ref{param_stars}). The mean velocity and metallicity of the cluster is RV=+54$\pm17$ km/s and $[Fe/H]=-0.48\pm0.1$, respectively. The results are tied to the measurements for Obj\,1, Obj\,2, and Obj\,4.  We estimated a reddening of $E(J-K)$=$2.0\pm0.3$ and distance modulus of $(M-m)_{0}$=$11.3\pm0.8$ (1.82 kpc). The age of this cluster candidate is approximately 1.3$\pm$0.4 Gyr.

\begin{figure}
\resizebox{8cm}{!}{\includegraphics{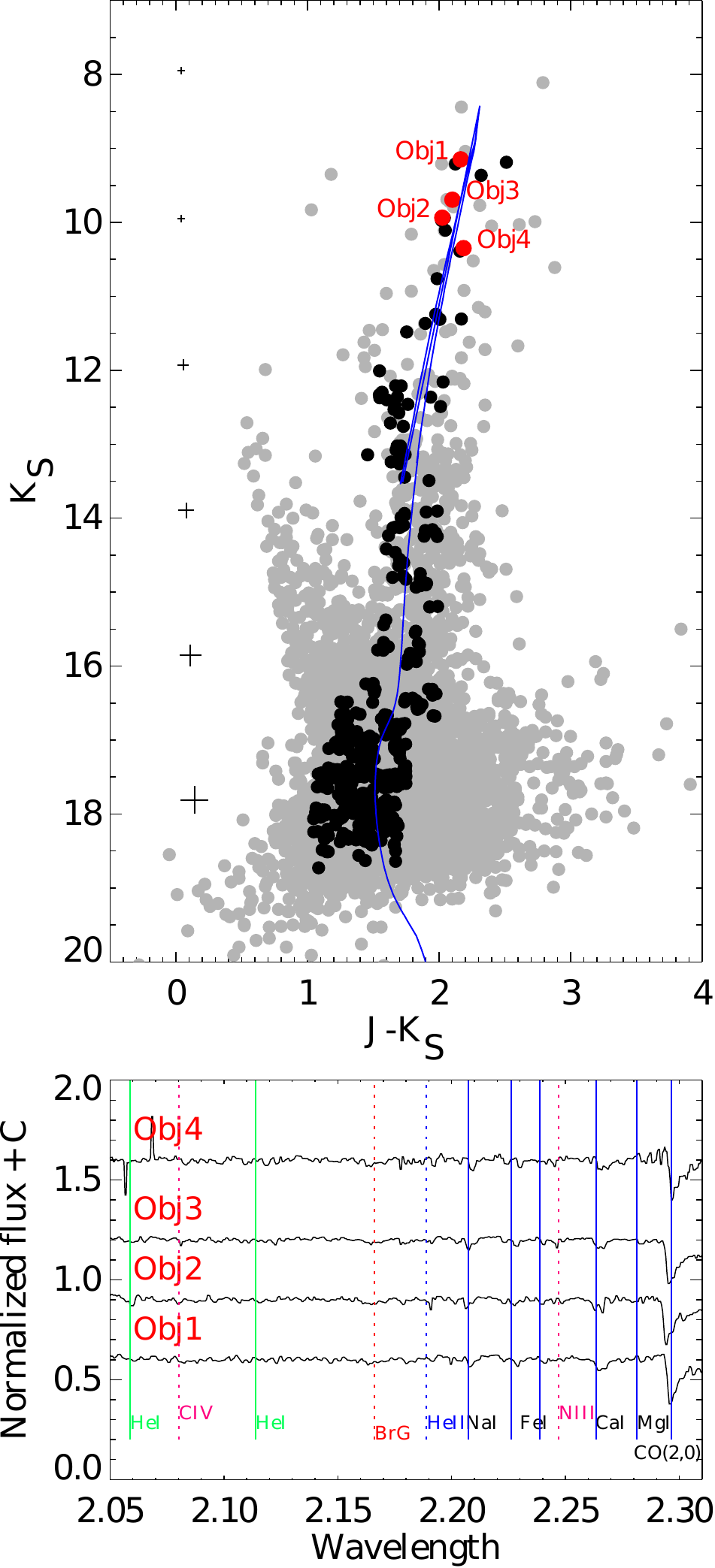}}
\caption{Top: $(J-K_{\rm S})$ vs. $K_{\rm S}$ color-magnitude diagram for CL\,150.  The symbols are the same as in Fig.~\ref{cl111_cmd}. The best fit is a 10 Gyr (z=0.003) Padova isochrone, which is shown. Bottom: SofI low resolution spectra of Obj\,1, Obj\,2, Obj\,3, and Obj\,4. }
\label{cl150_cmd}
\end{figure}

VVV\,CL150: The open cluster candidate lies in VVV tile b350 and appears confined within a radius of 60$"$. That makes this candidate one of the largest in our sample. Four stars were observed during our SofI run (Fig.~\ref{cl150_cmd}) and were subsequently classified as K2-5 giants. We performed PSF photometry using VVV-SkZ pipeline (Mauro et al. 2013) to construct the color-magnitude diagram. The diagram is well populated, and RC stars are readily discernible at $K_{\rm S}$=13.10$\pm0.2$ mag.  Moreover, the TO is likewise apparent at $K_{\rm S}$=17.2$\pm0.4$ mag. We estimated a reddening of $E(J-K)$=1.2$\pm0.1$, distance modulus of $(M-m)_{0}$=14.22$\pm0.7$ (6.98 kpc), and age of 10$\pm$0.8 Gyr.  CL\,150 is one of the most metal poor clusters in our sample, and it exhibits $[Fe/H]=-0.75\pm0.11$.  That determination relies on the metallicity measurements of all observed stars. Owing to its age and metallicity, CL\,150 is the most promising candidate for a new globular cluster in the Galactic bulge.
 
VVV\,CL152: The open cluster candidate lies in VVV tile b366. The candidate appears as a small stellar group confined within a radius of 35$"$. Eight stars were observed with SofI, and most are classified as G7-K0 giants.  Obj\,6 and Obj\,8 are O9-B0 and K3-4 V dwarfs, respectively. The VVV DR1 color-magnitude diagram displays evolved RGB stars and a relatively populated MS. We estimated a reddening of $E(J-K)$=1.4$\pm0.3$ and distance modulus of $(M-m)_{0}$=12.9$\pm0.4$ (3.8 kpc). The age of this candidate is estimated near 160$\pm$20 Myr. The metallicity of the cluster is $[Fe/H]=-0.90\pm0.37$, which is lower than expected for the estimated age. 
 
\begin{figure}
\resizebox{8cm}{!}{\includegraphics{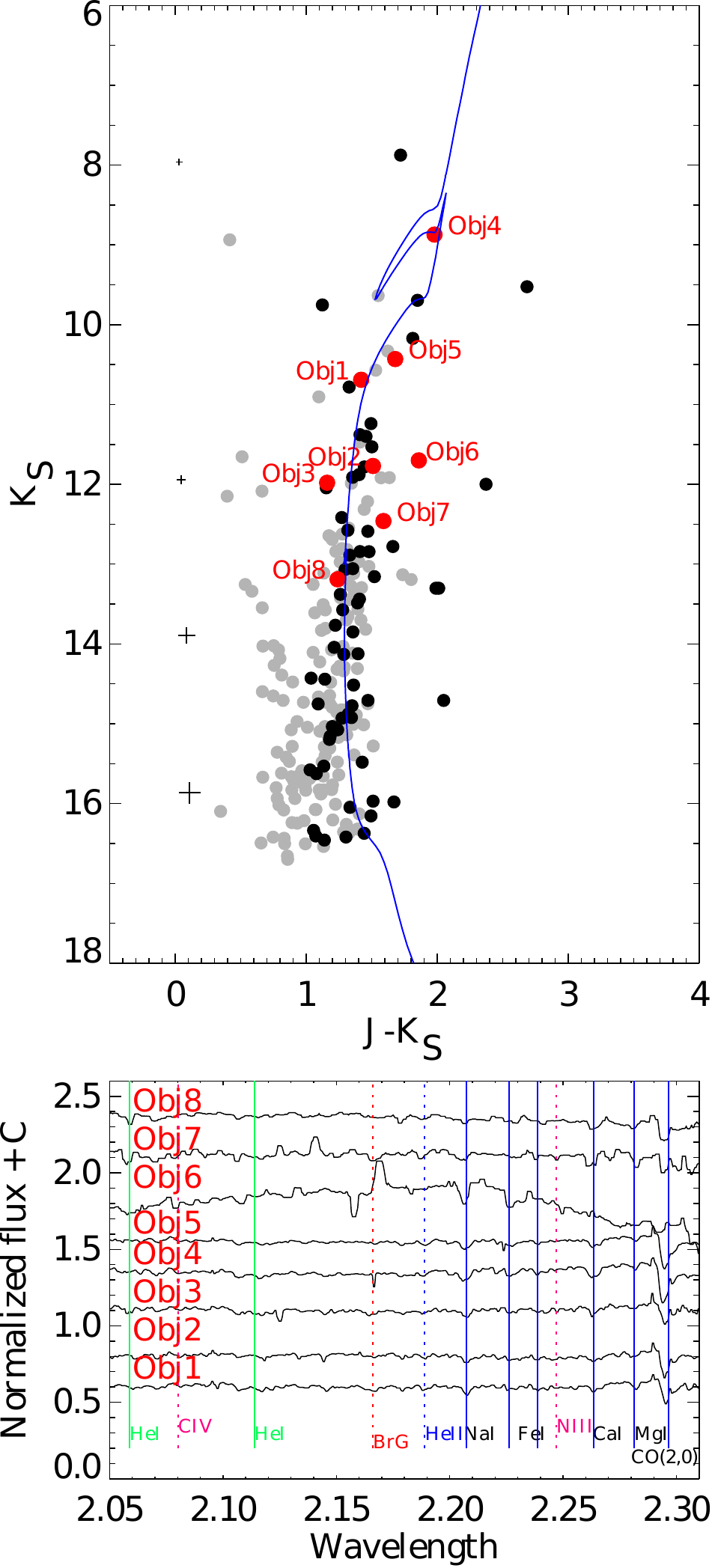}}
\caption{Top: $(J-K_{\rm S})$ vs. $K_{\rm S}$ color-magnitude diagram for CL\,152.  The symbols are the same as in Fig.~\ref{cl111_cmd}. The best fit is a 160 Myr (z=0.002) Padova isochrone, which is shown. Bottom: SofI low resolution spectra of observed objects. }
\label{cl152_cmd}
\end{figure}  


VVV\,CL154: The open cluster candidate lies in VVV tile b321 and appears as an overdensity confined within a radius of 30$"$. Three stars were observed during our SofI run and are classified as G5-K4 giants. The VVV DR1 color-magnitude diagram shows a poorly populated RGB, but a populated MS. We estimate a reddening of $E(J-K)$=1.2$\pm0.1$ and a  distance of $(M-m)_{0}$=10.1$\pm0.3$ (1.05 kpc). The age of this cluster candidate is approximately 8$\pm$0.2 Gyr. The metallicity of the cluster is $[Fe/H]=-0.69\pm0.4$, as inferred from measurements of Obj\,2 and Obj\,3 (Obj\,1 is rejected due to its position of the CMD). This is a good candidate for an old open cluster in close proximity to the Sun.  

\begin{figure}
\resizebox{8cm}{!}{\includegraphics{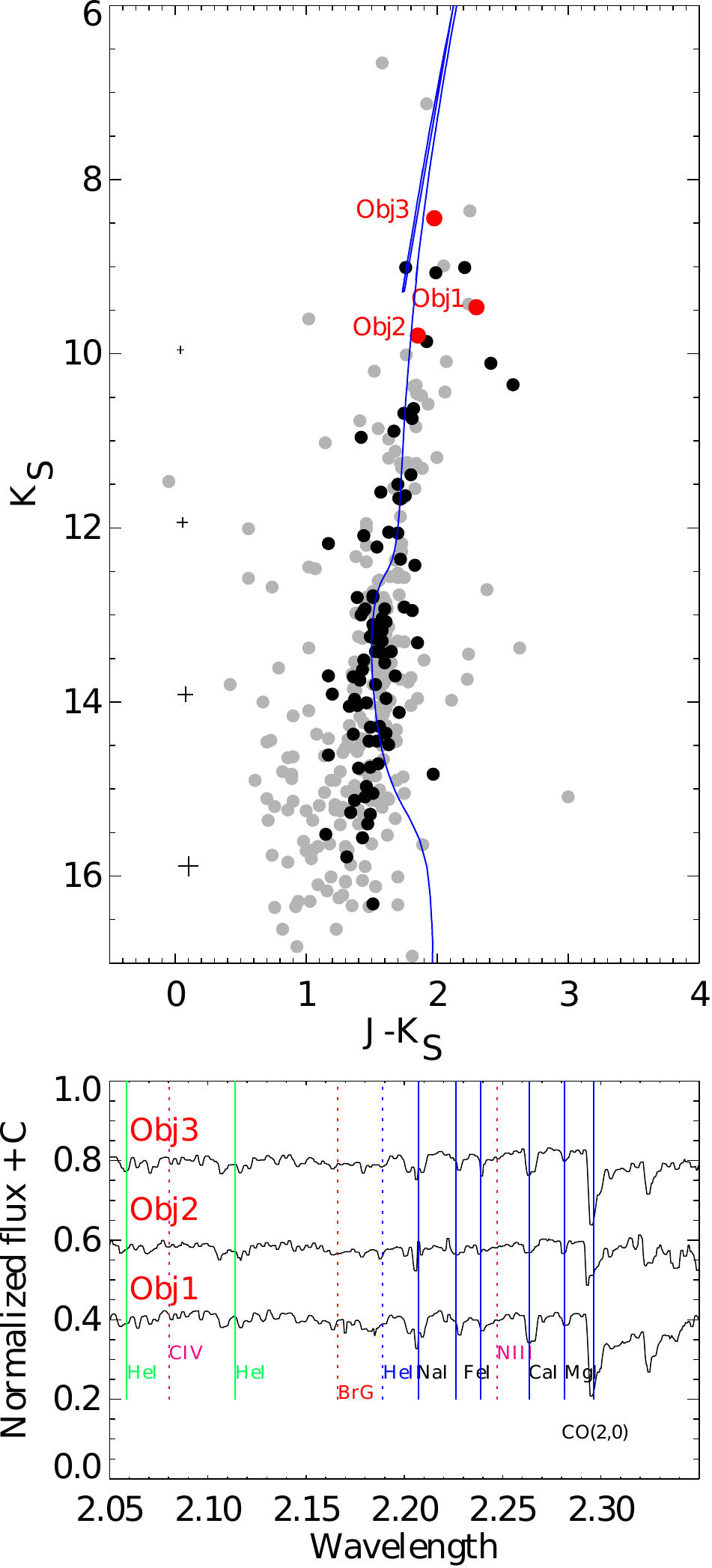}}
\caption{Top: $(J-K_{\rm S})$ vs. $K_{\rm S}$ color-magnitude diagram for CL\,154.  The symbols are the same as in  Fig.~\ref{cl111_cmd}. The best fit is an 8 Gyr (z=0.004) Padova isochrone, which is displayed. Bottom: SofI low resolution spectra of the observed objects.}
\label{cl154_cmd}
\end{figure}   

\begin{figure}
\resizebox{8cm}{!}{\includegraphics{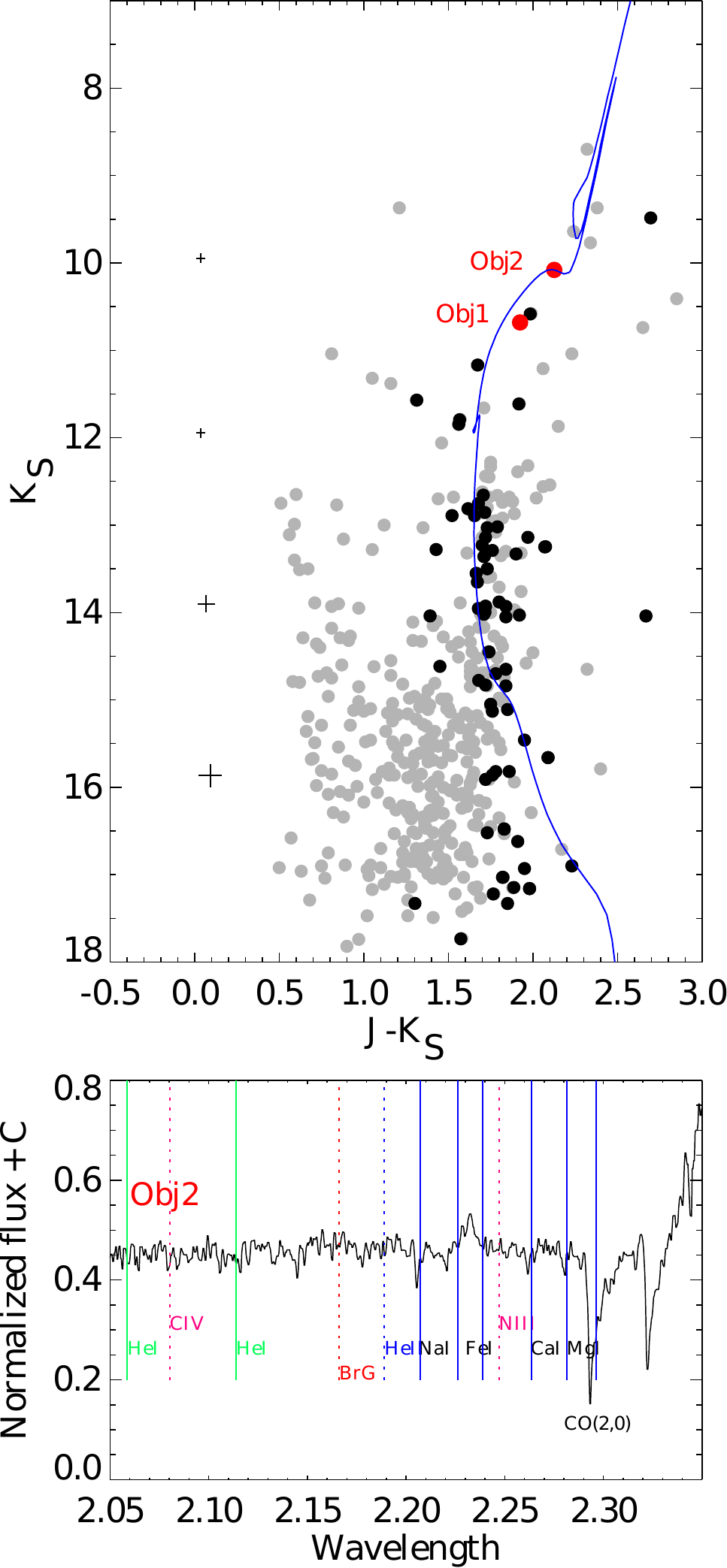}}
\caption{Top: $(J-K_{\rm S})$ vs. $K_{\rm S}$ color-magnitude diagram for CL\,157.  The symbols are the same as in  Fig.~\ref{cl111_cmd}. The best fit is a 316 Gyr (z=0.011) Padova isochrone, which is plotted. Bottom: SofI low resolution spectra of the observed objects.}
\label{cl157_cmd}
\end{figure}

VVV\,CL157: The open cluster candidate lies in VVV tile b354. It appears as an overdensity that is contained within a radius of 40$"$. Two stars were observed during our SofI run (Fig.~\ref{cl157_cmd}) and are classified as K2-3 giants. However, one of the stars exhibits low S/N and is not shown on the plot. The VVV DR1 database was used to construct the color-magnitude diagram, which shows few evolved stars and a poorly populated MS.  We estimated a reddening of $E(J-K)$=1.7$\pm0.1$ and distance modulus of $(M-m)_{0}$=11.7$\pm0.2$ (2.19 kpc). The age and metallicity of this cluster candidate is approximately 316$\pm$38 Myr and  $[Fe/H]=-0.23\pm0.15$, respectively.  The latter estimate relies solely on Obj\,2.

\begin{figure}
\resizebox{8cm}{!}{\includegraphics{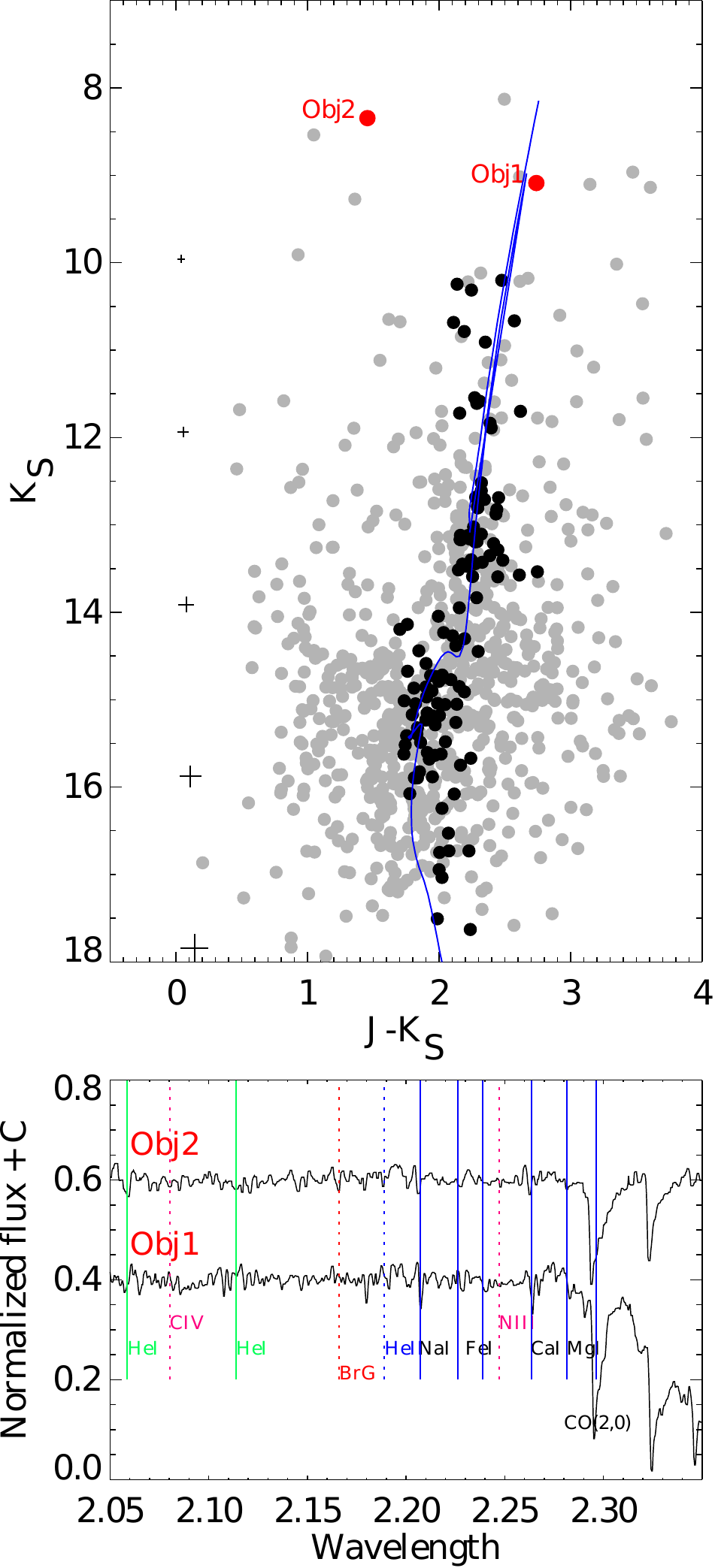}}
\caption{Top: $(J-K_{\rm S})$ vs. $K_{\rm S}$ color-magnitude diagram for CL\,160.  The symbols are the same as in  Fig.~\ref{cl111_cmd}. The best fit is a 1.6 Gyr (z=0.004) Padova isochrone, which is shown. Bottom: SofI low resolution spectra of the observed objects.}
\label{cl160_cmd}
\end{figure}

VVV\,CL160: The open cluster candidate lies in VVV tile b340 and appears as an overdensity that occupies a radius of 55$"$. Two stars were observed during our SOAR run (Fig.~\ref{cl160_cmd}) and are classified as K0-4 giants. The VVV DR1 database was used to construct the color-magnitude diagram, which exhibits a defined RGB and MS. Object 2 lies far from the cluster center and features a lower velocity than Obj\,1. The Besan\c{c}on model predicts RV=+45$\pm$75 km/s in this direction.  The former star likely belongs to the field. The RC can be identified at $K_{\rm S}$=13.1$\pm0.3$ mag, while the TO point appears at $K_{\rm S}$=14.8$\pm0.4$ mag. We estimated a reddening of $E(J-K)$=1.72$\pm0.1$ and distance modulus of $(M-m)_{0}$=13.60$\pm0.3$ (5.25 kpc). The age of this star cluster candidate is approximately 1.6$\pm$0.5 Gyr. The cluster metallicity is $[Fe/H]=-0.72\pm0.21$, but the result relies solely on Obj\,1. In sum, this candidate is probably an old metal-poor open cluster.

\begin{figure}
\resizebox{8cm}{!}{\includegraphics{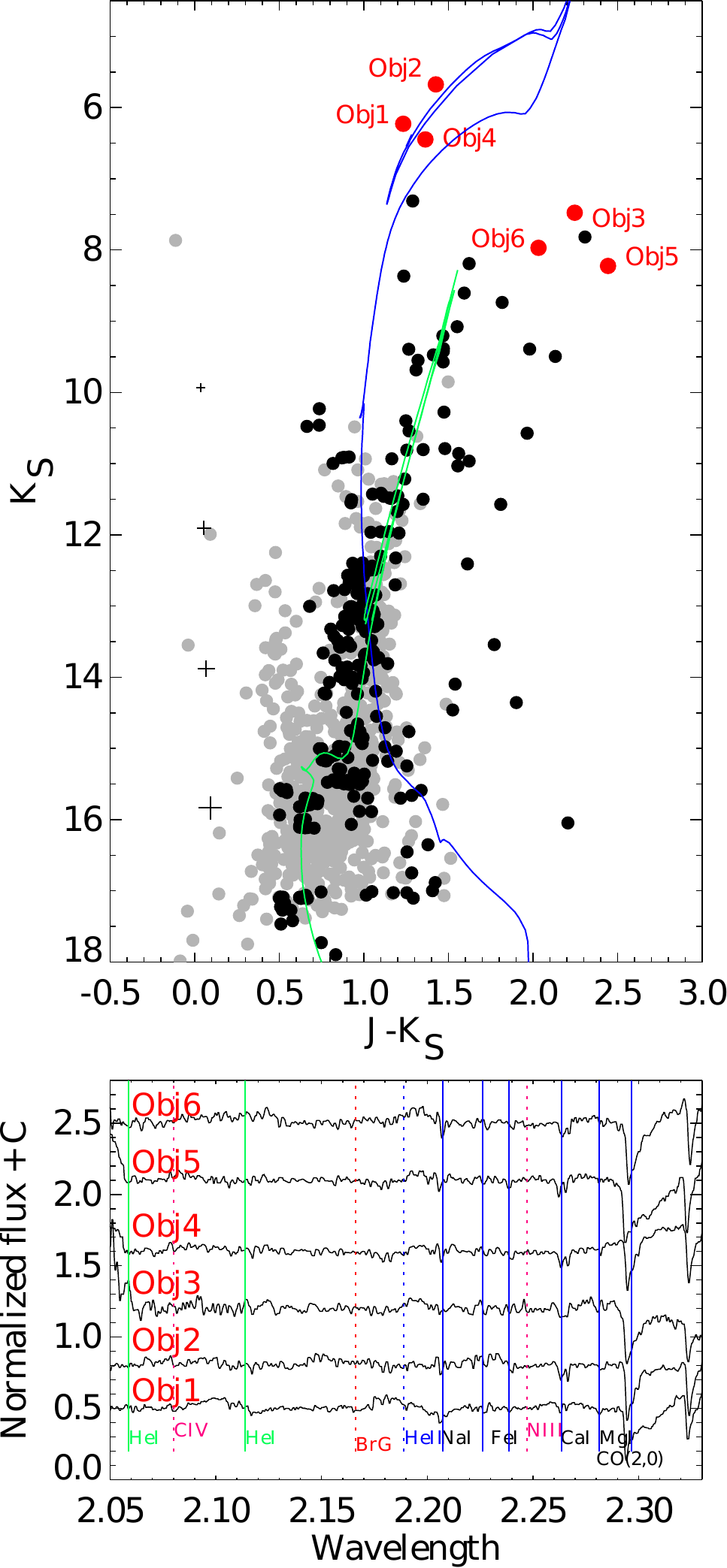}}
\caption{Top: $(J-K_{\rm S})$ vs. $K_{\rm S}$ color-magnitude diagram for CL\,161.  The symbols are the same as in  Fig.~\ref{cl111_cmd}. Both 25 Myr (blue) and 2 Gyr (green) Padova isochrones (z=0.018) are shown (see text).  Bottom: SofI low resolution spectra of the observed objects.}
\label{cl161_cmd}
\end{figure}

VVV\,CL161: The open cluster candidate lies in VVV tile b295. Ten bright stars form an overdensity that spans a radius of 75$"$. Six of those stars were observed with SOAR during the 2012 run (Fig.~\ref{cl161_cmd}). The stars can be classified as K3-M5 giants or K0-M1 supergiants.  The low S/N and resolution of the spectra give rise to that degeneracy.  In the VVV DR1 color-magnitude diagram those bright stars form two groups and suffer from different extinction. According to the Messineo et al. (2012) Q1 and Q2 parameters, the stars designated Obj\,1, Obj\,2, and Obj\,4 could be red supergiants. Adopting the RSG cluster hypothesis, we derived a reddening of $E(J-K)$=1.15$\pm0.12$ and distance modulus of $(M-m)_{0}$=12.70$\pm0.3$ (3.47 kpc).  The cluster age is approximately 25 $\pm$6 Myr.  Obj\,3, Obj\,5, and Obj\,6 are field stars under the RSG cluster scenario. Alternatively, if stars near $K_{\rm S}$=13.2$\pm0.2$ mag are of the RG class, and the TO lies near $K_{\rm S}$=16.0$\pm0.4$ mag, the cluster parameters are $E(J-K)$=0.4$\pm0.2$, $(M-m)_{0}$=14.50$\pm0.4$ (7.94 kpc), and  2$\pm$0.5 Gyr. Deeper photometry and higher resolution spectroscopy are needed to verify the nature of this candidate.

The parameters derived for the 20 clusters described above are listed in Table~\ref{param_clusters}\footnote{Table~\ref{param_clusters} is available in electronic form at the CDS, either via anonymous ftp to cdsarc.u-strasbg.fr (130.79.128.5) or http://cdsweb.u-strasbg.fr/cgi-bin/qcat?J/A+A/}. The column headers are as follows: name of the cluster, mean color excess of the cluster, distance in kpc, age from the isochrone fit in Myr or Gyr, mean metallicity of the cluster ([Fe/H] in dex), mean radial velocity RV in km/s, and the reddening derived from Gonzalez et al. (2011, 2012) as a comparison. For CL\,130 and CL\,160, we give the derived parameters for both red-supergiant and red-giant spectral classifications of the observed stars.  

In Table~\ref{param_stars}\footnote{Table~\ref{param_stars} is available in electronic form at the CDS, either via anonymous ftp to cdsarc.u-strasbg.fr (130.79.128.5) or http://cdsweb.u-strasbg.fr/cgi-bin/qcat?J/A+A/} we highlight the parameters derived for individual cluster stars that possess spectra. The column headers are as follows: names of the cluster and constituent star, coordinates in degrees, near-infrared photometric (J, H, and $K_{S}$) magnitudes, the assigned spectral type, individual reddening derived from its spectral type, the individual radial velocities in km/s, individual abundance estimates ([Fe/H] in dex), and date and instrument tied to the observations.

\section{Summary} 

In this paper we have reported the discovery of 58 star cluster candidates projected on the Galactic bulge and inner disk area.  The candidates were identified in near-infrared images and photometry associated with the ``VVV -- Vista Variables in the V\'{\i}a L\'actea'' ESO Large Survey. Relative to the  Borissova et al. (2011) cluster sample, the catalog presented here contains older candidates that feature larger angular radii. Most clusters in the present sample are not embedded or near nebulosity.  However, the candidates detected are highly obscured and suffer from upwards of $E(J-K)=5.6$\,mag of extinction. For 20 cluster we have the fundamental parameters such as reddening,  age, distance, and metallicity.  The most interesting candidates are as follows. The color-magnitude diagrams of  VVV CL\,119, VVV CL\,143, and VVV CL\,150  show well-defined red giant branch stars and some red clump and main sequence stars. They are projected at 6.8, 7.8, and 6.98 kpc, are 2-10 Gyr old and intermediate metal poor, and could be classified as old open clusters. However, these objects in the inner few kpc from the Galactic center are quite unusual, because they should be rare in the inner Galaxy. Thus, these are promising candidates for new globular clusters in the galactic bulge. The cluster candidates VVV CL\,139 and VVV CL\,140 are projected very close to each other and show similar radial velocities and distance modulus of 3.8 kpc. The age of CL139  is estimated around 80 Myr, while CL140 is older (1.3 Gyr).  Both clusters are relatively metal rich and are good new cluster pair-candidates. And finally, three cluster candidates from our sample, namely VVV CL\,117, VVV CL\,130 and VVV CL\,161 show typical color-magnitude diagrams of red supergiant clusters, but more data are needed to confirm their nature.

\begin{acknowledgements}
The data used in this paper were obtained with VIRCAM/VISTA at the ESO Paranal Observatory. The VVV Survey is supported by BASAL Center for Astrophysics and Associated Technologies CATA PFB-06, by the Ministry of Economy, Development, and Tourism’s Millennium Science Initiative through grant IC12009, awarded to The Millennium Institute of Astrophysics (MAS). This publication makes use of data products from the Two Micron All Sky Survey, which is a joint project of the University of Massachusetts and the Infrared Processing and Analysis Center/California Institute of Technology, funded by the National Aeronautics and Space Administration and the National Science Foundation. This research has made use of the  Aladin  and SIMBAD database, operated at the CDS, Strasbourg, France. Support  for JB and RK  is provided from Fondecyt Reg. No. 1120601 and  No. 1130140 and Cento de Astrof\'isica de Valpara\'iso. DM is supported by FONDECYT No. 1130196. CB and EB acknowledge support from Brazil's CNPq.  MG is financed by the GEMINI-CONICYT Fund, allocated to Project 32110014. The SOAR Telescope is a joint project of: Conselho Nacional de Pesquisas Cientificas e Tecnoligicas CNPq-Brazil, The University of North Carolina at Chapel Hill, Michigan State University, and the National Optical Astronomy Observatory. IRAF is distributed by the National Optical Astronomy Observatory, which is operated by the Association of Universities for Research in Astronomy (AURA) under cooperative agreement with the National Science Foundation. IN is partially supported by  Gemini-Conycit No.32120017.  MMH and BP acknowledge support by the National Science Foundation under Grant No. 1009550. IN and AM acknowledge support for this work by the Spanish Government Ministerio de Ciencia e Innovaci\'on (MICINN) through grant AYA2012-39364-C02-02. R.K.S. acknowledges support from CNPq/Brazil through projects 310636/2013-2 and 481468/2013-7. SRA was supported by the FONDECYT project No. 3140605. P. Amigo is supported by the ALMA-CONICYT project number 31110002. S.R.A. was supported by the FONDECYT project No. 3140605. We thank the anonymous referee for the useful suggestions that helped to improve our paper.

\end{acknowledgements}

\newpage

\onecolumn

\begin{longtable}{lcccccl}\small\\
\caption{VVV cluster candidates.}
\label{candidates}\\
\hline\hline
\multicolumn{1}{l}{Name} &
\multicolumn{1}{c}{RA(J2000)} &
\multicolumn{1}{c}{DEC(J2000)}&
\multicolumn{1}{c}{Tile} &
\multicolumn{1}{c}{Radius} &
\multicolumn{1}{c}{Numb.} &
\multicolumn{1}{l}{Comments }\small\\
&hh:mm:ss&deg:mm:ss& &arcsec &   \\
\endfirsthead
\caption{continued.}\\
\hline\hline
\multicolumn{1}{l}{Name} &
\multicolumn{1}{c}{RA(J2000)} &
\multicolumn{1}{c}{DEC(J2000)}&
\multicolumn{1}{c}{Tile} &
\multicolumn{1}{c}{Radius} &
\multicolumn{1}{c}{Numb.} &
\multicolumn{1}{l}{Comments }\small\\
&hh:mm:ss&deg:mm:ss& &arcsec &   \\
\hline
\endhead
\hline
\endfoot   
\hline
VVV CL106	&	17:17:09	&	-36:22:00	&	b341	&	20	&	10	&\tiny	close to DBSB 121 and DBSB 122	\\
VVV CL107	&	17:17:27	&	-36:14:44	&	b341	&	30	&	10	&\tiny	6-7 bright blue stars, stellar group	\\
VVV CL108	&	17:19:22	&	-34:16:31	&	b342	&	30	&	45	&\tiny	concentrated, has a core	\\
VVV CL109	&	17:21:36	&	-35:32:52	&	b328	&	31	&	30	&\tiny	close to [CWP2007] CS78 bubble, at edge of dark nebula	\\
VVV CL110	&	17:22:47	&	-34:41:17	&	b342	&	20	&	30	&\tiny	GlC candidate or Old OC?	\\
VVV CL111	&	17:22:50	&	-36:34:16	&	b327	&	35	&	58	&\tiny	old open cluster	\\
VVV CL112	&	17:24:14	&	-32:45:30	&	b343	&	30	&	40	&\tiny	loose cluster	\\
VVV CL113	&	17:24:34	&	-33:18:35	&	d343	&	20	&	23	&\tiny	compact group	\\
VVV CL114	&	17:25:13	&	-37:07:38	&	b313	&	12	&	10	&\tiny	nebulosity, group	\\
VVV CL115	&	17:26:23	&	-34:34:53	&	b329	&	45	&	30	&\tiny	at edge of dark nebula, OC candidate	\\
VVV CL116	&	17:27:31	&	-35:16:28	&	b328	&	20	&	10	&\tiny	nebulosity,Young Cluster, Dark Nebula around	\\
VVV CL117	&	17:30:13	&	-34:02:37	&	b329	&	12	&	20	&\tiny	red supergiant candidate	\\
VVV CL118	&	17:30:35	&	-35:08:02	&	b315	&	20	&	37	&\tiny	bright stars, low overdensity	\\
VVV CL119	&	17:30:46	&	-32:39:05	&	b330	&	55	&	157	&\tiny	old open cluster	\\
VVV CL120	&	17:32:21	&	-36:25:48	&	b300	&	25	&	72	&\tiny	old open cluster	\\
VVV CL121	&	17:32:22	&	-35:26:22	&	b315	&	34	&	50	&\tiny	bright stars, low overdensity	\\
VVV CL122	&	17:35:46	&	-31:54:39	&	b331	&	30	&	27	&\tiny	nebulosity,[CWP2007] CS28, CS30, in dark nebula\\
VVV CL123	&	17:36:27	&	-35:39:16	&	b301	&	40	&	67	&\tiny	old open cluster	\\
VVV CL124	&	17:36:35	&	-29:34:54	&	b346	&	15	&	15	&\tiny	open cluster	\\
VVV CL125	&	17:34:29	&	-30:27:09	&	b346	&	30	&	16  &\tiny  bright stars, overdensity  	\\
VVV CL126	&	17:38:06	&	-31:30:34	&	b331	&	40	&	44	&\tiny	loose OC 	\\
VVV CL127	&	17:38:46	&	-28:29:57	&	b347	&	25	&	33	&\tiny	loose cluster?   	\\
VVV CL128	&	17:39:59	&	-32:26:27	&	b317	&	30	&	100	&\tiny	GlC candidate or Old OC	\\
VVV CL129	&	17:40:44	&	-30:09:25	&	b332	&	20	&	19	&\tiny	several red stars, loose cluster	\\
VVV CL130	&	17:41:11	&	-30:26:31	&	b332	&	20	&	24	&\tiny	red supergiant candidate	\\
VVV CL131	&	17:41:17	&	-34:34:02	&	b302	&	50	&	87	&\tiny	old OC or GlC candidate 	\\
VVV CL132	&	17:41:37	&	-29:44:08	&	b333	&	20	&	20	&\tiny	compact cluster	\\
VVV CL133	&	17:41:43	&	-29:44:46	&	b333	&	16	&	20	&\tiny	compact cluster	\\
VVV CL134	&	17:42:30	&	-30:01:17	&	b332	&	12	&	14	&\tiny	embedded cluster	\\
VVV CL135	&	17:42:43	&	-29:51:35	&	b333	&	30	&	15	&\tiny	embedded cluster or dust window	\\
VVV CL136	&	17:42:57	&	-29:52:40	&	b333	&	10	&	12	&\tiny	compact cluster   	\\
VVV CL137	&	17:43:01	&	-31:23:12	&	b318	&	45	&	70	&\tiny	Young Cluster or Dust Window?	\\
VVV CL138	&	17:43:40	&	-30:27:43	&	b318	&	10	&	25	&\tiny	Compact, Dark Nebulae around	\\
VVV CL139	&	17:43:50	&	-31:44:16	&	b318	&	20	&	47	&\tiny	open cluster, pair with CL140	\\
VVV CL140	&	17:43:53	&	-31:43:59	&	b318	&	20	&	51	&\tiny	open cluster, pair with CL139	\\
VVV CL141	&	17:44:05	&	-27:56:10	&	b348	&	10	&	16	&\tiny	embedded cluster	\\
VVV CL142	&	17:44:34	&	-28:39:52	&	b333	&	20	&	31	&\tiny	compact open cluster or edge of dark nebula	\\
VVV CL143	&	17:44:36	&	-33:44:18	&	b302	&	50	&	93	&\tiny	old open cluster or young globular cluster	\\
VVV CL144	&	17:45:35	&	-25:28:14	&	b363	&	10	&	12	&\tiny	compact cluster	\\
VVV CL145	&	17:45:38	&	-25:27:38	&	b363	&	25	&	38	&\tiny	poorly populated OC	\\
VVV CL146	&	17:46:00	&	-28:49:09	&	b333	&	20	&	49	&\tiny	open cluster or dust window	\\
VVV CL147	&	17:46:28	&	-28:39:18	&	b334	&	10	&	5	  &\tiny	nebulosity, Maser,IR, in dark cloud	\\
VVV CL148	&	17:47:20	&	-29:11:55	&	b319	&	10	&	5	  &\tiny	embedded 	\\
VVV CL149	&	17:49:22	&	-29:27:56	&	b319	&	20	&	41	&\tiny	open cluster	\\
VVV CL150	&	17:50:41	&	-25:13:06	&	b350	&	25	&	312	&\tiny	old open cluster or young globular cluster	\\
VVV CL151	&	17:51:17	&	-29:39:04	&	b319	&	30	&	38	&\tiny	prominent OC 	\\
VVV CL152	&	17:53:08	&	-22:38:41	&	b366	&	20	&	64	&\tiny	open cluster	\\
VVV CL153	&	17:53:32	&	-25:22:56	&	b336	&	30	&	56	&\tiny	prominent bright cluster	\\
VVV CL154	&	17:55:08	&	-28:06:01	&	b321	&	20	&	38	&\tiny	old open cluster	\\
VVV CL155	&	18:01:00	&	-23:46:12	&	b338	&	15	&	12	&\tiny	concentrated coordinates a bit off\\
VVV CL156	&	18:02:12	&	-22:07:34	&	b339	&	30	&	49	&\tiny	big cluster or field	\\
VVV CL157	&	18:04:06	&	-19:43:50	&	b354	&	35	&	39	&\tiny	old open cluster	\\
VVV CL158	&	18:05:34	&	-21:52:58	&	b339	&	25	&	31	&\tiny	nebulosity, close to bubbles, embedded dark nebula around	\\
VVV CL159	&	18:06:27	&	-21:38:57	&	b339	&	10	&	15	&\tiny	faint, very reddened, in dark cloud, embedded	\\
VVV CL160	&	18:06:57	&	-20:00:40	&	b340	&	25	&	114	&\tiny	old open cluster	\\
VVV CL161	&	18:07:40	&	-26:11:51	&	b295	&	75	&	10	&\tiny	overdensity	\\
VVV CL162	&	18:09:42	&	-21:46:56	&	b325	&	20	&	27	&\tiny	OC	\\
VVV CL163	&	18:13:57	&	-20:42:07	&	b326	&	15	&	24	&\tiny	rather loose OC	\\
\hline
\end{longtable}	 							

\newpage

\begin{longtable}{llllllc}\small\\
\caption{Parameters for the relatively sizable VVV cluster candidates.}\\
\label{param_clusters}\\
\hline\hline
\multicolumn{1}{l}{Name} &
\multicolumn{1}{c}{E$(J-K_{\rm S})$} &
\multicolumn{1}{c}{Dist.} &
\multicolumn{1}{c}{Age} &
\multicolumn{1}{c}{[Fe/H]} &
\multicolumn{1}{c}{RV} &
\multicolumn{1}{c}{ E$(J-K_{\rm S})$} \\
      &mag&kpc& &dex&km/s&G2011  \\
\endfirsthead
\caption{continued.}\\
\hline\hline
\multicolumn{1}{l}{Name} &
\multicolumn{1}{c}{E$(J-K_{\rm S})$} &
\multicolumn{1}{c}{Dist.} &
\multicolumn{1}{c}{Age} &
\multicolumn{1}{c}{[Fe/H]} &
\multicolumn{1}{c}{RV} &
\multicolumn{1}{c}{E$(J-K_{\rm S})$Gon} \\
      &mag&kpc& &dex&km/s&mag  \\
\hline
\endhead
\hline
\endfoot   
\hline
VVV	CL111						&	3.4$\pm$0.4		&	4.17$\pm$0.8	&	1.6$\pm$0.7\,Gyr	&	$+$0.43$\pm$0.20	&	$-$77$\pm$21	&	3.09	\\
VVV	CL113						&2.3$\pm$0.4		&	5.25$\pm$1		&	32$\pm$7\,Myr			&	$-$0.23$\pm$0.15	&	$+$7$\pm$23		&	1.56	\\
VVV	CL117						&	2.9$\pm$0.6		&	11.75$\pm$1		&	20$\pm$5\,Myr			&										&	$+$72$\pm$25	&	2.22	\\
VVV	CL119						&	2.03$\pm$0.4	&	6.8$\pm$0.7		&	5$\pm$1.2\,Gyr		&	$-$0.30$\pm$0.18	&	$+$96$\pm$29	&	2.15	\\
VVV	CL120						&	1.2$\pm$0.1		&	2.09$\pm$0.4	&	2$\pm$0.5\,Gyr		&	$-$0.36$\pm$0.45	&	$+$51$\pm$9		&	0.87	\\
VVV	CL123						&	0.55$\pm$0.2	&	2.63$\pm$0.4	&	9.0$\pm$0.7\,Gyr	&	$-$0.19$\pm$0.14	&	$-$50$\pm$17	&	0.55	\\
VVV	CL124						&	1.8$\pm$0.3		&	5.25$\pm$0.8	&	50$\pm$6\,Myr			&	$-$0.40$\pm$0.50	&	$-$36$\pm$8		&	0.98	\\
VVV	CL130\footnote{Red Supergiants} &	5.6$\pm$0.3		&	1.58$\pm$0.2	&	32$\pm$8\,Myr			&										&	$-$20$\pm$8		&	3.40	\\
VVV	CL130\footnote{Red Giants}		&	5.2$\pm$0.3		&	1.9$\pm$0.3		&	316$\pm$37\,Myr		&										&	$-$20$\pm$8		&	3.40	\\
VVV	CL139						&	2.6$\pm$0.3		&	3.8$\pm$0.7		&	80$\pm$19\,Myr		&	$-$0.20$\pm$0.25	&	$-$35$\pm$25	&	1.92	\\
VVV	CL140						&	2.3$\pm$0.4		&	3.8$\pm$0.7		&	1.3$\pm$0.3\,Gyr	&	$-$0.42$\pm$0.35	&	$-$19$\pm$4		&	1.92	\\
VVV	CL142						&	4.8$\pm$0.3		&	1.8$\pm$0.3		&	800$\pm$92\,Myr		&	$-$0.14$\pm$0.40	&	$+$14$\pm$17	&	3.66	\\
VVV	CL143						&	0.58$\pm$0.2	&	7.8$\pm$1.5		&	4$\pm$0.7\,Gyr		&	$-$0.62$\pm$0.52	&	$+$86$\pm$26	&	0.45	\\
VVV	CL146						&	5.0$\pm$0.5		&	4.37$\pm$0.8	&	40$\pm$7\,Myr			&	$-$0.15$\pm$0.25	&								&	4.20	\\
VVV	CL149						&	2.0$\pm$0.3		&	1.82$\pm$0.5	&	1.3$\pm$0.4\,Gyr	&	$-$0.48$\pm$0.1		&	$+$54$\pm$17	&	1.43	\\
VVV	CL150						&	1.2$\pm$0.1		&	6.98$\pm$1.2	&	10$\pm$0.8\,Gyr		&	$-$0.75$\pm$0.11	&								&	1.20	\\
VVV	CL152						&	1.4$\pm$0.3		&	3.8$\pm$0.5		&	160$\pm$20\,Myr		&	$-$0.90$\pm$0.37	&								&	0.70	\\
VVV	CL154						&	1.2$\pm$0.1		&	1.05$\pm$0.1	&	8$\pm$0.2\,Gyr		&	$-$0.69$\pm$0.4		&								&	0.88	\\
VVV	CL157						&	1.7$\pm$0.1		&	2.19$\pm$0.1	&	316$\pm$38\,Myr		&	$-$0.23$\pm$0.15	&								&	1.20	\\
VVV	CL160						&	1.72$\pm$0.1	&	5.25$\pm$0.5	&	1.6$\pm$0.5\,Gyr	&	$-$0.72$\pm$0.21	&								&	1.73	\\
VVV	CL161\footnote{Red Supergiants}	&	1.15$\pm$0.12	&	3.47$\pm$0.3	&	25$\pm$0.6\,Myr		&										&								&	0.43	\\
VVV	CL161\footnote{Red Giants}	&	0.4$\pm$0.2		&	7.94$\pm$1		&	2$\pm$0.5\,Gyr		&										&								&	0.43	\\
\hline					
\end{longtable}


\tiny
\begin{landscape}
\begin{longtable}{cllrrrl@{\hspace{-3pt}}l@{\hspace{-3pt}}rll}
\caption{Parameters for the individual cluster stars.}
\label{param_stars}\\
\hline\hline
\multicolumn{1}{c}{Name} & 
\multicolumn{1}{c}{RA} & 
\multicolumn{1}{c}{DEC}& 
\multicolumn{1}{c}{J} &  
\multicolumn{1}{c}{H} & 
\multicolumn{1}{c}{K$_{\rm S}$} & 
\multicolumn{1}{c}{Sp.Type}& 
\multicolumn{1}{c}{E$(J$$-$$K_{\rm S})$} & 
\multicolumn{1}{l}{RV} & 
\multicolumn{1}{c}{[Fe/H]} & 
\multicolumn{1}{l}{Log}\\
  & \multicolumn{1}{c}{deg} & \multicolumn{1}{c}{deg} & \multicolumn{1}{c}{mag} & \multicolumn{1}{c}{mag} & \multicolumn{1}{c}{mag} & & \multicolumn{1}{c}{mag}  &\multicolumn{1}{l}{km/s}  & \multicolumn{1}{c}{dex} \\
\endfirsthead
\caption{continued.}\\
\hline\hline
\multicolumn{1}{c}{Name} &
\multicolumn{1}{c}{RA} &
\multicolumn{1}{c}{DEC}&
\multicolumn{1}{c}{J} &
\multicolumn{1}{c}{H} &
\multicolumn{1}{c}{K$_{\rm S}$} &
\multicolumn{1}{c}{Sp.Type}&
\multicolumn{1}{l}{E$(J$$-$$K_{\rm S})$} &
\multicolumn{1}{l}{RV} &
\multicolumn{1}{c}{[Fe/H]} &
\multicolumn{1}{l}{Log}\\
 & \multicolumn{1}{c}{deg} & \multicolumn{1}{c}{deg} & \multicolumn{1}{c}{mag} & \multicolumn{1}{c}{mag} & \multicolumn{1}{c}{mag} & & \multicolumn{1}{c}{mag}  &\multicolumn{1}{l}{km/s}  & \multicolumn{1}{c}{dex} \\
\hline
\endhead
\hline
\endfoot   
\hline		
CL111Obj3	&	260.709080	&	-36.571393	&	16.74$\pm$0.03	&	13.67$\pm$0.03	&	12.55$\pm$0.02	&	K4-5III	&	3.00$\pm$0.12	&								&											&	2012-05-04,SofI	\\
CL111Obj4	&	260.708095	&	-36.571335	&	15.18$\pm$0.10	&	12.18$\pm$0.06	&	10.61$\pm$0.04	&	K0-1III	&	4.02$\pm$0.12	&	$-$77$\pm$23	&	$+$0.43$\pm$0.15		&	2012-05-04,SofI	\\
CL113Obj1	&	261.139858	&	-33.309269	&	9.77$\pm$0.04		&	7.9$\pm$0.06		&	6.95$\pm$0.02		&	M0-1III	&	2.59$\pm$0.07	&	 		 					&	$-$0.16$\pm$0.12		&	2012-05-04,SofI	\\
CL113Obj2	&	261.139260	&	-33.313648	&	11.64$\pm$0.04	&	9.62$\pm$0.03		&	8.71$\pm$0.03		&	M0-1III	&	2.70$\pm$0.07	&	$+$7$\pm$24		&	$-$0.30$\pm$0.14		&	2012-05-04,SofI	\\
CL117Obj1	&	262.556760	&	-34.044170	&	12.14$\pm$0.04	&	9.78$\pm$0.05		&	8.34$\pm$0.06		&	K0-1I		&	2.92$\pm$0.09	&	$+$63$\pm$34	&	 	 	 								&	2012-05-04,SofI	\\
CL117Obj2	&	262.555067	&	-34.044415	&	11.69$\pm$0.06	&	9.39$\pm$0.08		&	8.16$\pm$0.04		&	G8-9I		&	2.74$\pm$0.05	&	90$\pm$22			&	 	 	 								&	2012-05-04,SofI	\\
CL117Obj3	&	262.556359	&	-34.042732	&	13.06$\pm$0.06	&	10.55$\pm$0.06	&	9.28$\pm$0.03		&	K0-1I		&	2.91$\pm$0.09	&	32$\pm$21			&	 	 	 								&	2012-05-04,SofI	\\
CL117Obj4	&	262.565930	&	-34.038769	&	12.98$\pm$0.06	&	10.61$\pm$0.06	&	9.49$\pm$0.05		&	K0-1I		&	2.00$\pm$0.09	&	60$\pm$17			&	 	 	 								&	2012-05-04,SofI	\\
CL117Obj5	&	262.550472	&	-34.044296	&	16.62$\pm$0.04	&	14.52$\pm$0.02	&	13.76$\pm$0.02	&	late	star&							&	10$\pm$27			&											&	2012-05-04,SofI	\\
CL119Obj1	&	262.694627	&	-32.645966	&	8.26$\pm$0.02		&	6.74$\pm$0.05		&	5.98$\pm$0.03		&	K0-1III	&	1.88$\pm$0.12	&	108$\pm$21		&	$-$0.07$\pm$0.11		&	2012-05-04,SofI	\\
CL119Obj2	&	262.692057	&	-32.657436	&	9.49$\pm$0.02		&	7.89$\pm$0.06		&	7.13$\pm$0.02		&	K3-4III	&	1.81$\pm$0.15	&	107$\pm$27		&	$-$0.25$\pm$0.13		&	2012-05-04,SofI	\\
CL119Obj3	&	262.691191	&	-32.645840	&	13.53$\pm$0.15	&	11.18$\pm$0.06	&	10.36$\pm$0.02	&	K3-4III	&	2.68$\pm$0.15	&	79$\pm$34			&	$-$0.31$\pm$0.14		&	2012-05-04,SofI	\\
CL119Obj4	&	262.688083	&	-32.648609	&	13.09$\pm$0.02	&	11.03$\pm$0.02	&	10.11$\pm$0.03	&	K2-3III	&	2.48$\pm$0.16	&	112$\pm$23		&	$-$0.36$\pm$0.17		&	2012-05-04,SofI	\\
CL119Obj5	&	262.690626	&	-32.646135	&	14.7$\pm$0.01		&	12.55$\pm$0.01	&	11.66$\pm$0.01	&	K2-3III	&	2.55$\pm$0.16	&	-7$\pm$16			&	$-$0.60$\pm$0.21		&	2012-05-04,SofI	\\
CL119Obj6	&	262.685225	&	-32.651222	&	12.15$\pm$0.03	&	9.95$\pm$0.03		&	9.0$\pm$0.03		&	K1-2III	&	2.7$\pm$0.08	&	153$\pm$33		&	$-$0.52$\pm$0.24		&	2012-05-04,SofI	\\
CL120Obj1	&	263.088451	&	-36.428932	&	10.48$\pm$0.04	&	8.91$\pm$0.02 	&	8.13$\pm$0.02		&	M6-7III	&	1.64$\pm$0.06	&								&	 	 	 								&	2012-05-03,SofI	\\
CL120Obj2	&	263.089948	&	-36.430817	&	9.16$\pm$0.02		&	7.7$\pm$0.04		&	7.02$\pm$0.03		&	M1-2III	&	1.49$\pm$0.09	&	54$\pm$16			&	$-$0.07$\pm$0.10		&	2012-05-03,SofI	\\
CL120Obj3	&	263.087256	&	-36.423359	&	11.55$\pm$0.03	&	10.09$\pm$0.03	&	9.55$\pm$0.02		&	K1-2III	&	1.56$\pm$0.08	&	48$\pm$22			&	$-$0.66$\pm$0.21		&	2012-05-03,SofI	\\
CL120Obj4	&	263.085037	&	-36.429344	&	9.8$\pm$0.08		&	8.69$\pm$0.04		&	8.98$\pm$0.04		&	K4-5III	&	0.16$\pm$0.12	&	111$\pm$27		&	$-$0.12$\pm$0.11		&	2012-05-03,SofI	\\
CL120Obj5	&	263.084764	&	-36.430126	&	10.02$\pm$0.05	&	8.57$\pm$0.02		&	8.53$\pm$0.06		&	K3-4III	&	0.84$\pm$0.15	&	152$\pm$20		&	$-$0.12$\pm$0.11		&	2012-05-03,SofI	\\
Cl123Obj1	&	264.120148	&	-35.656879	&	9.95$\pm$0.02		&	8.81$\pm$0.03		&	8.35$\pm$0.02		&	K1-2III	&	1.15$\pm$0.08	&	-36$\pm$9			&	$-$0.03$\pm$0.12		&	2012-05-03,SofI	\\
Cl123Obj2	&	264.112250	&	-35.651253	&	11.68$\pm$0.04	&	10.66$\pm$0.04	&	10.26$\pm$0.02	&	K1-2III	&	0.97$\pm$0.08	&	 				 			&	$-$0.16$\pm$0.14		&	2012-05-03,SofI	\\
CL123Obj3	&	264.107672	&	-35.647839	&	8.16$\pm$0.02		&	7.1$\pm$0.02		&	6.76$\pm$0.02		&	M3			&	0.74$\pm$0.16	&	-46$\pm$13		&	$-$0.25$\pm$0.13		&	2012-05-03,SofI	\\
CL123Obj4	&	264.110832	&	-35.660934	&	7.99$\pm$0.03		&	6.84$\pm$0.03		&	6.32$\pm$0.03		&	M6			&	1.03$\pm$0.12	&	-67$\pm$10		&	$-$0.31$\pm$0.15		&	2012-05-03,SofI	\\
CL123Obj5	&	264.104189	&	-35.654114	&	9.27$\pm$0.03		&	8.51$\pm$0.02		&	8.25$\pm$0.03		&	G8-9III	&	0.65$\pm$0.07	&	6	$\pm$	21		&	$-$0.26$\pm$0.11		&	2012-05-03,SofI	\\
CL124Obj1	&	264.145178	&	-29.580191	&	11.44$\pm$0.07	&	10.02$\pm$0.10	&	9.3$\pm$0.06		&	K3-4III	&	1.59$\pm$0.15	&	-37$\pm$16		&	$-$0.03$\pm$0.10		&	2012-05-06,SofI	\\
CL124Obj2	&	264.144743	&	-29.581400	&	10.46$\pm$0.05	&	10.41$\pm$0.08	&	8.45$\pm$0.05		&	K4-5III	&	1.41$\pm$0.12	&	-46$\pm$19		&	$-$0.01$\pm$0.11		&	2012-05-06,SofI	\\
CL124Obj3	&	264.143778	&	-29.582207	&	10.54$\pm$0.06	&	9.57$\pm$0.10		&	8.86$\pm$0.04		&	K1-2III	&	1.24$\pm$0.08	&	-39$\pm$25		&	$-$0.93$\pm$0.21		&	2012-05-06,SofI	\\
CL124Obj4	&	264.145985	&	-29.581848	&	11.08$\pm$0.07	&	9.52$\pm$0.09		&	8.51$\pm$0.04		&	K0-1III	&	2.14$\pm$0.12	&	-30$\pm$26		&	$-$1.03$\pm$0.34		&	2012-05-06,SofI	\\
CL124Obj5	&	264.145014	&	-29.582500	&	10.54$\pm$0.05	&	10.26$\pm$0.08	&	8.51$\pm$0.06		&	M1-2III	&	1.36$\pm$0.09	&	-27$\pm$19	  &	$+$0.05$\pm$0.10 	 	&	2012-05-06,SofI	\\
CL130Obj1	&	265.295189	&	-30.440063	&	12.76$\pm$0.03	&	9.08$\pm$0.03		&	7.22$\pm$0.02		&	M0-1I		&	4.52$\pm$0.05	&	-15$\pm$31	  &	$+$0.08$\pm$0.12		&	2012-05-03,SofI	\\
CL130Obj2	&	265.293421	&	-30.441717	&	12.01$\pm$0.02	&	8.55$\pm$0.04		&	6.92$\pm$0.02		&	K3-4I		&	4.28$\pm$0.10	&	-42$\pm$35	  &	$+$0.62$\pm$0.24		&	2012-05-03,SofI	\\
CL130Obj3	&	265.301513	&	-30.437988	&	13.73$\pm$0.03	&	10.65$\pm$0.02	&	9.27$\pm$0.02		&	K1-2III	&	4.01$\pm$0.08	&	 							&	$+$0.10$\pm$0.14		&	2012-05-03,SofI	\\
CL130Obj4	&	265.297311	&	-30.441914	&	13.15$\pm$0.06	&	8.73$\pm$0.04		&	6.46$\pm$0.02		&	K3-4I		&	5.88$\pm$0.10	&	-16$\pm$27		&											&	2012-05-03,SofI	\\
CL130Obj5	&	265.296287	&	-30.442829	&	13.32$\pm$0.04	&	9.64$\pm$0.03		&	7.23$\pm$0.08		&	M0-1I		&	5.08$\pm$0.05	&	18$\pm$13			&	$-$0.02$\pm$0.11		&	2012-05-03,SofI	\\
CL139Obj1	&	265.957411	&	-31.737024	&	13.05$\pm$0.06	&	10.89$\pm$0.09	&	9.59$\pm$0.05		&	M0-1III	&	2.79$\pm$0.07	&	 						 	&	$-$0.39$\pm$0.14		&	2012-05-06,SofI	\\
CL139Obj2	&	265.955805	&	-31.739756	&	11.82$\pm$0.06	&	10.11$\pm$0.03	&	9.15$\pm$0.03		&	M0-1III	&	2.00$\pm$0.07	&	-35$\pm$25		&	$-$0.01$\pm$0.11		&	2012-05-06,SofI	\\
CL140Obj1	&	265.971981	&	-31.737429	&	11.63$\pm$0.02	&	9.12$\pm$0.03		&	7.91$\pm$0.02		&	M1III		&	3.04$\pm$0.07	&							 	&	$-$0.10$\pm$0.13		&	2012-05-06,SofI	\\
CL140Obj2	&	265.974657	&	-31.736923	&	11.90$\pm$0.02	&	9.63$\pm$0.03		&	8.65$\pm$0.02		&	K4-5III	&	2.66$\pm$0.12	&	-21$\pm$28		&	$-$0.42$\pm$0.15		&	2012-05-06,SofI	\\
CL140Obj3	&	265.973223	&	-31.731306	&	14.43$\pm$0.05	&	11.67$\pm$0.04	&	10.4$\pm$0.04		&	G9III		&	3.66$\pm$0.05	&	87$\pm$19			&	$-$1.02$\pm$0.23		&	2012-05-06,SofI	\\
CL140Obj4	&	265.971785	&	-31.730059	&	12.90$\pm$0.04	&	10.67$\pm$0.06	&	9.69$\pm$0.05		&	K3III		&	2.72$\pm$0.14	&	-16$\pm$39		&	$-$0.74$\pm$0.19		&	2012-05-06,SofI	\\
CL140Obj5	&	265.970468	&	-31.729139	&	12.31$\pm$0.05	&	10.77$\pm$0.03	&	10.11$\pm$0.03	&	K0III		&	1.80$\pm$0.08	&	2$\pm$33			&	$-$0.11$\pm$0.14		&	2012-05-06,SofI	\\
CL142Obj1	&	266.143009	&	-28.663368	&	13.81$\pm$0.07	&	10.66$\pm$0.08	&	8.84$\pm$0.07		&	K0III		&	4.58$\pm$0.08	&	45$\pm$30			&	$-$0.86$\pm$0.22		&	2012-05-02,SofI	\\
CL142Obj2	&	266.142651	&	-28.664537	&	14.84$\pm$0.08	&	11.23$\pm$0.07	&	9.42$\pm$0.07		&	M5-8III	&	4.78$\pm$0.08	&		 						&	$+$0.33$\pm$0.12		&	2012-05-02,SofI	\\
CL142Obj3	&	266.142397	&	-28.665148	&	16.71$\pm$0.03	&	12.63$\pm$0.05	&	10.88$\pm$0.05	&	M0III		&	5.21$\pm$0.05	&		 						&	$-$0.03$\pm$0.11		&	2012-05-02,SofI	\\
CL142Obj4	&	266.141444	&	-28.664568	&	14.78$\pm$0.11	&	10.92$\pm$0.08	&	8.80$\pm$0.08		&	M3-4III	&	5.32$\pm$0.10	&	17$\pm$23			&	$+$0.05$\pm$0.11		&	2012-05-02,SofI	\\
CL142Obj5	&	266.143705	&	-28.664934	&	15.36$\pm$0.07	&	11.58$\pm$0.04	&	9.69$\pm$0.04		&	M0III		&	5.06$\pm$0.05	&	12$\pm$25			&	$-$0.17$\pm$0.12		&	2012-05-02,SofI	\\
CL143Obj1	&	266.153022	&	-33.740978	&	10.79$\pm$0.03	&	9.6$\pm$0.03		&	9.15$\pm$0.03		&	M0-1III	&	0.98$\pm$0.07	&	102$\pm$23		&	$+$0.02$\pm$0.12		&	2012-05-02,SofI	\\
CL143Obj2	&	266.147631	&	-33.735596	&	10.57$\pm$0.04	&	9.46$\pm$0.04		&	9.15$\pm$0.04		&	M0III		&	0.80$\pm$0.05	&	79$\pm$15			&	$-$1.13$\pm$0.32		&	2012-05-02,SofI	\\
CL143Obj3	&	266.146035	&	-33.733501	&	11.42$\pm$0.04	&	10.38$\pm$0.03	&	10.05$\pm$0.03	&	K0III		&	0.97$\pm$0.08	&	119$\pm$18		&	$-$0.99$\pm$0.21		&	2012-05-02,SofI	\\
CL143Obj4	&	266.165291	&	-33.733170	&	10.74$\pm$0.03	&	9.5$\pm$0.03		&	8.97$\pm$0.03		&	M0III		&	1.16$\pm$0.05	&	 		 					&	$-$1.17$\pm$0.27		&	2012-05-02,SofI	\\
CL143Obj5	&	266.156442	&	-33.736031	&	10.35$\pm$0.03	&	9.19$\pm$0.03		&	8.63$\pm$0.03		&	M0III		&	1.10$\pm$0.05	&	88$\pm$21			&	$-$0.16$\pm$0.11		&	2012-05-02,SofI	\\
CL143Obj6	&	266.145125	&	-33.739616	&	11.13$\pm$0.03	&	9.92$\pm$0.03		&	9.59$\pm$0.04		&	K3III		&	1.35$\pm$0.14	&	60$\pm$18			&	$-$0.25$\pm$0.12		&	2012-05-02,SofI	\\
CL143Obj7	&	266.154177	&	-33.736938	&	10.77$\pm$0.03	&	9.65$\pm$0.03		&	9.36$\pm$0.03		&	G8III		&	1.04$\pm$0.04	&	67$\pm$28			&	$-$1.05$\pm$0.25		&	2012-05-02,SofI	\\
CL146Obj1	&	266.503501	&	-28.817527	&	15.22$\pm$0.08	&	11.33$\pm$0.05	&	9.35$\pm$0.05		&	M5III		&	4.88$\pm$0.06	&								&											&	2012-05-14,OSIRIS	\\
CL146Obj2	&	266.500850	&	-28.817808	&	15.11$\pm$0.15	&	11.53$\pm$0.15	&	9.36$\pm$0.05		&	M5-7III	&	5.12$\pm$0.08	&							 	&	$-$0.01$\pm$0.12		&	2012-05-14,OSIRIS	\\
CL146Obj3	&	266.499904	&	-28.818137	&	15.13$\pm$0.01	&	12.05$\pm$0.01	&	10.55$\pm$0.01	&	K5III		&	3.97$\pm$0.09	&	114$\pm$21		&	$-$0.37$\pm$0.17		&	2012-05-14,OSIRIS	\\
CL146Obj4	&	266.498021	&	-28.818344	&	13.57$\pm$0.04	&	9.83$\pm$0.05		&	8.06$\pm$0.03		&	K5I			&	4.55$\pm$0.06	&	133$\pm$21		&	$+$0.15$\pm$0.11		&	2012-05-14,OSIRIS	\\
CL146Obj5	&	266.496767	&	-28.818642	&	14.25$\pm$0.05	&	12.14$\pm$0.07	&	8.93$\pm$0.05		&	M0III		&	4.70$\pm$0.05	&	121$\pm$29		&	$-$0.36$\pm$0.15		&	2012-05-14,OSIRIS	\\
CL149Obj1	&	267.347058	&	-29.462183	&	13.90$\pm$0.10	&	12.29$\pm$0.08	&	11.53$\pm$0.07	&	K1III		&	1.95$\pm$0.07	&	54$\pm$23			&	$-$0.50$\pm$0.16		&	2012-05-02,SofI	\\
CL149Obj2	&	267.345025	&	-29.464018	&	12.27$\pm$0.07	&	10.44$\pm$0.07	&	9.60$\pm$0.04		&	M0III		&	2.05$\pm$0.05	&	53$\pm$23			&	$-$0.53$\pm$0.15		&	2012-05-02,SofI	\\
CL149Obj3	&	267.343054	&	-29.466053	&	7.58$\pm$0.02		&	6.62$\pm$0.04		&	6.25$\pm$0.02		&	G9-K0III&	0.93$\pm$0.09	&	-91$\pm$29		&	$+$0.02$\pm$0.11		&	2012-05-02,SofI	\\
CL149Obj4	&	267.340996	&	-29.467297	&	10.52$\pm$0.05	&	8.6$\pm$0.03		&	7.55$\pm$0.02		&	K4-5III	&	2.42$\pm$0.12	&	55$\pm$15			&	$-$0.40$\pm$0.12		&	2012-05-02,SofI	\\
CL149Obj5	&	267.338429	&	-29.463346	&	9.09$\pm$0.06		&	7.52$\pm$0.10		&	6.90$\pm$0.04		&	K4-5III	&	1.64$\pm$0.12	&	-41$\pm$18		&	$-$0.30$\pm$0.21		&	2012-05-02,SofI	\\
CL149Obj6	&	267.338266	&	-29.464169	&	8.65$\pm$0.03		&	7.36$\pm$0.03		&	6.85$\pm$0.02		&	K2-3III	&	1.35$\pm$0.16	&	36$\pm$30			&											&	2012-05-02,SofI	\\
CL150Obj1	&	267.677058	&	-25.218052	&	11.31$\pm$0.03	&	9.79$\pm$0.03		&	9.15$\pm$0.02		&	K3III		&	1.68$\pm$0.14	&	 	 	 					&	$-$0.83$\pm$0.13		&	2012-05-05,SofI	\\
CL150Obj2	&	267.670466	&	-25.217829	&	11.96$\pm$0.05	&	10.52$\pm$0.04	&	9.94$\pm$0.05		&	K2-3III	&	1.58$\pm$0.16	&	 	 	 					&	$-$0.65$\pm$0.11		&	2012-05-05,SofI	\\
CL150Obj3	&	267.668962	&	-25.217709	&	11.80$\pm$0.07	&	10.29$\pm$0.07	&	9.69$\pm$0.06		&	K4-5III	&	1.55$\pm$0.12	&	 	 	 					&	$-$0.77$\pm$0.17		&	2012-05-05,SofI	\\
CL150Obj4	&	267.665514	&	-25.217760	&	12.54$\pm$0.04	&	10.99$\pm$0.04	&	10.35$\pm$0.03	&	K4-5III	&	1.64$\pm$0.12	&	50$\pm$10			&	$-$0.76$\pm$0.15		&	2012-05-05,SofI	\\
CL152Obj1	&	268.281935	&	-22.643370	&	12.11$\pm$0.08	&	11.16$\pm$0.10	&	10.69$\pm$0.07	&	G9III		&	1.05$\pm$0.05	&	 	 	 					&	$-$1.12$\pm$0.21		&	2011-04-19,SofI	\\
CL152Obj2	&	268.282059	&	-22.643873	&	13.28$\pm$0.01	&	12.22$\pm$0.01	&	11.77$\pm$0.01	&	G9III		&	1.14$\pm$0.05	&	 	 	 					&	$-$1.09$\pm$0.27		&	2011-04-19,SofI	\\
CL152Obj3	&	268.282080	&	-22.644624	&	13.14$\pm$0.01	&	12.22$\pm$0.01	&	11.98$\pm$0.01	&	G9III		&	0.79$\pm$0.05	&	 	 	 					&	$-$0.96$\pm$0.22		&	2011-04-19,SofI	\\
CL152Obj4	&	268.281984	&	-22.646515	&	10.85$\pm$0.05	&	9.45$\pm$0.05		&	8.87$\pm$0.04		&	G7III		&	1.60$\pm$0.04	&	 	 	 					&	$-$1.11$\pm$0.19		&	2011-04-19,SofI	\\
CL152Obj5	&	268.282065	&	-22.647198	&	12.11$\pm$0.01	&	11.36$\pm$0.01	&	10.43$\pm$0.01	&	K0-3III	&	1.23$\pm$0.19	&	 	 	 					&	$-$0.32$\pm$0.15		&	2011-04-19,SofI	\\
CL152Obj6	&	268.282567	&	-22.652021	&	13.56$\pm$0.06	&	12.20$\pm$0.06	&	11.7$\pm$0.05		&	O9V			&	2.03$\pm$0.04	&	 		 					&											&	2011-04-19,SofI	\\
CL152Obj7	&	268.282738	&	-22.653069	&	14.05$\pm$0.07	&	12.88$\pm$0.07	&	12.46$\pm$0.05	&	G8-K0III&	1.20$\pm$0.06	&	 	 	 					&	$-$0.80$\pm$0.18		&	2011-04-19,SofI	\\
CL152Obj8	&	268.282276	&	-22.656670	&	14.43$\pm$0.09	&	13.44$\pm$0.09	&	13.19$\pm$0.07	&	K3V			&	0.59$\pm$0.08	&	 		 					&											&	2011-04-19,SofI	\\
CL154Obj1	&	268.786860	&	-28.099880	&	11.77$\pm$0.04	&	10.22$\pm$0.04	&	9.47$\pm$0.03		&	K0-4III	&	1.85$\pm$0.19	&	 	 	 					&	$-$0.19$\pm$0.11		&	2011-04-19,SofI	\\
CL154Obj2	&	268.783761	&	-28.100582	&	11.64$\pm$0.04	&	10.40$\pm$0.05	&	9.79$\pm$0.04		&	G5III		&	1.50$\pm$0.32	&	 	 	 					&	$-$0.45$\pm$0.13		&	2011-04-19,SofI	\\
CL154Obj3	&	268.780811	&	-28.101231	&	10.42$\pm$0.01	&	9.06$\pm$0.01		&	8.44$\pm$0.06		&	K0-4III	&	1.53$\pm$0.19	&	 	 	 					&	$-$0.93$\pm$0.27		&	2011-04-19,SofI	\\
CL157Obj1	&	271.022438	&	-19.729368	&	12.61$\pm$0.03	&	11.25$\pm$0.03	&	10.68$\pm$0.03	&	K2III		&	1.47$\pm$0.12	&	 		 					&											&	2012-05-14,OSIRIS	\\
CL157Obj2	&	271.024838	&	-19.733290	&	12.21$\pm$0.04	&	10.71$\pm$0.04	&	10.08$\pm$0.03	&	K2-3III	&	1.64$\pm$0.16	&	-14$\pm$16		&	$-$0.23$\pm$0.15		&	2012-05-14,OSIRIS	\\
CL160Obj1	&	271.740450	&	-20.007586	&	11.83$\pm$0.03	&	9.85$\pm$0.03		&	9.09$\pm$0.03		&	K2-4III	&	2.29$\pm$0.17	&	185$\pm$10		&	$-$0.72$\pm$0.21		&	2012-05-14,OSIRIS	\\
CL160Obj2	&	271.736140	&	-20.009516	&	9.80$\pm$0.02		&	8.79$\pm$0.04		&	8.35$\pm$0.03		&	K0III		&	1.06$\pm$0.08	&	20$\pm$14			&	 										&	2012-05-14,OSIRIS	\\
CL161Obj1	&	271.912188	&	-26.207048	&	7.46$\pm$0.03		&	6.51$\pm$0.03		&	6.23$\pm$0.02		&	K0-1I		&	0.68$\pm$0.09	&	 		 					&											&	2012-05-14,OSIRIS	\\
CL161Obj2	&	271.910090	&	-26.196815	&	7.10$\pm$0.02		&	6.12$\pm$0.02		&	5.68$\pm$0.02		&	K5-6I		&	0.88$\pm$0.07	&	 		 					&											&	2012-05-14,OSIRIS	\\
CL161Obj3	&	271.924521	&	-26.192663	&	9.72$\pm$0.02		&	8.32$\pm$0.03		&	7.48$\pm$0.03		&	M0-1I		&	1.61$\pm$0.05	&	 		 					&											&	2012-05-14,OSIRIS	\\
CL161Obj4	&	271.914693	&	-26.190865	&	7.81$\pm$0.02		&	6.82$\pm$0.04		&	6.45$\pm$0.02		&	K8-9I		&	0.81$\pm$0.06	&	 		 					&											&	2012-05-14,OSIRIS	\\
CL161Obj5	&	271.930827	&	-26.207802	&	10.67$\pm$0.03	&	8.61$\pm$0.03		&	8.22$\pm$0.03		&	K1-2I		&	1.80$\pm$0.08	&	 		 					&											&	2012-05-14,OSIRIS	\\
CL161Obj6	&	271.930011	&	-26.207237	&	10.00$\pm$0.05	&	8.48$\pm$0.03		&	7.97$\pm$0.02		&	K2-3I		&	1.39$\pm$0.12	&	 		 					&											&	2012-05-14,OSIRIS	\\
\hline					
\end{longtable}	
\end{landscape}																																																																																																																																																																							
\newpage

\appendix

\vspace{5cm}
\onecolumn{}  

\begin{figure}[h]
\resizebox{17cm}{!}{\includegraphics{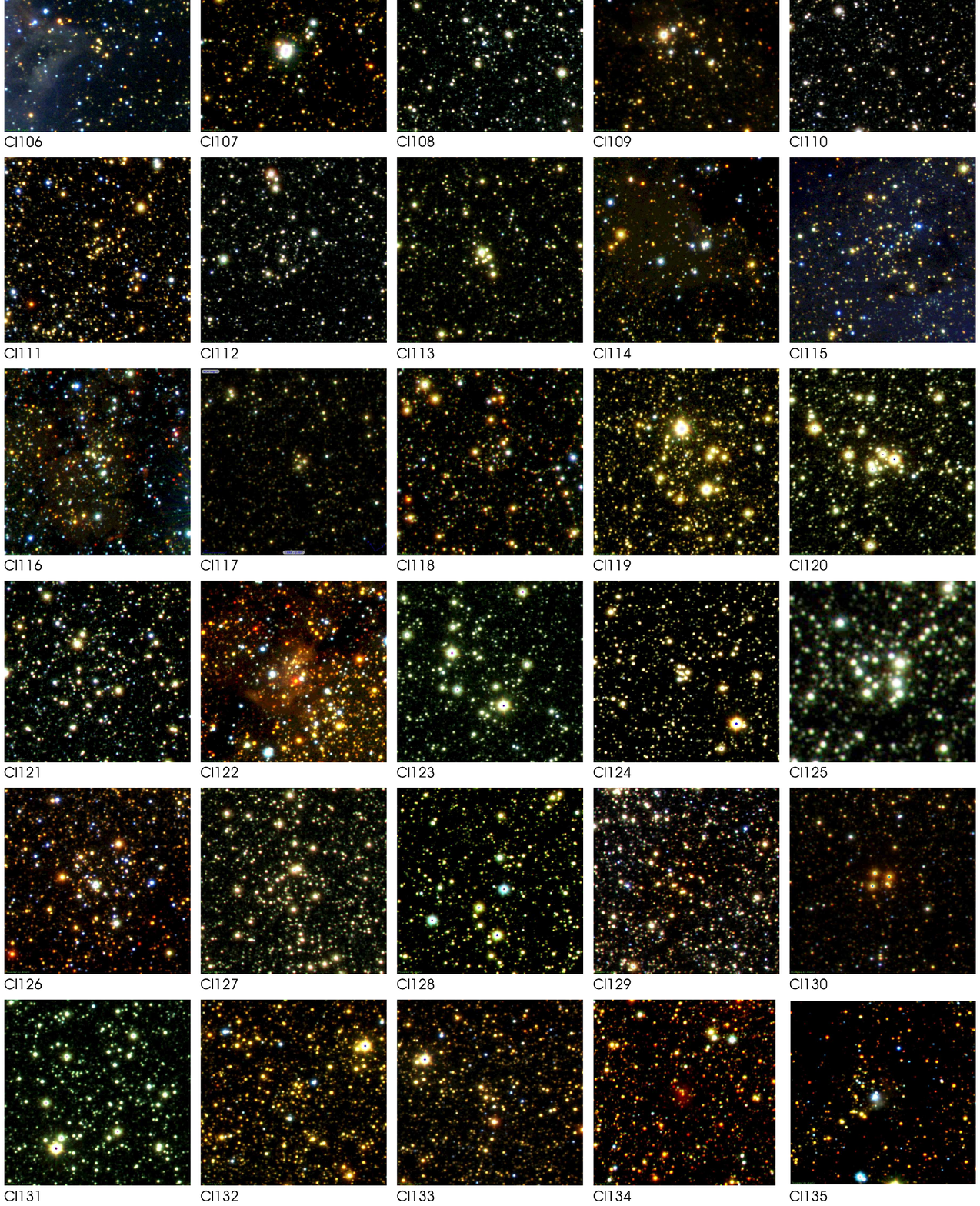}}
\caption{VVV $JHK_{\rm S}$ composite color images of VVV open cluster candidates. The field of view is approx. 2.5$\times$2.5 arcmin, and north is up, east to the right.}
\end{figure}
 
\newpage
\onecolumn{}  
\begin{figure}[h]
\resizebox{17cm}{!}{\includegraphics{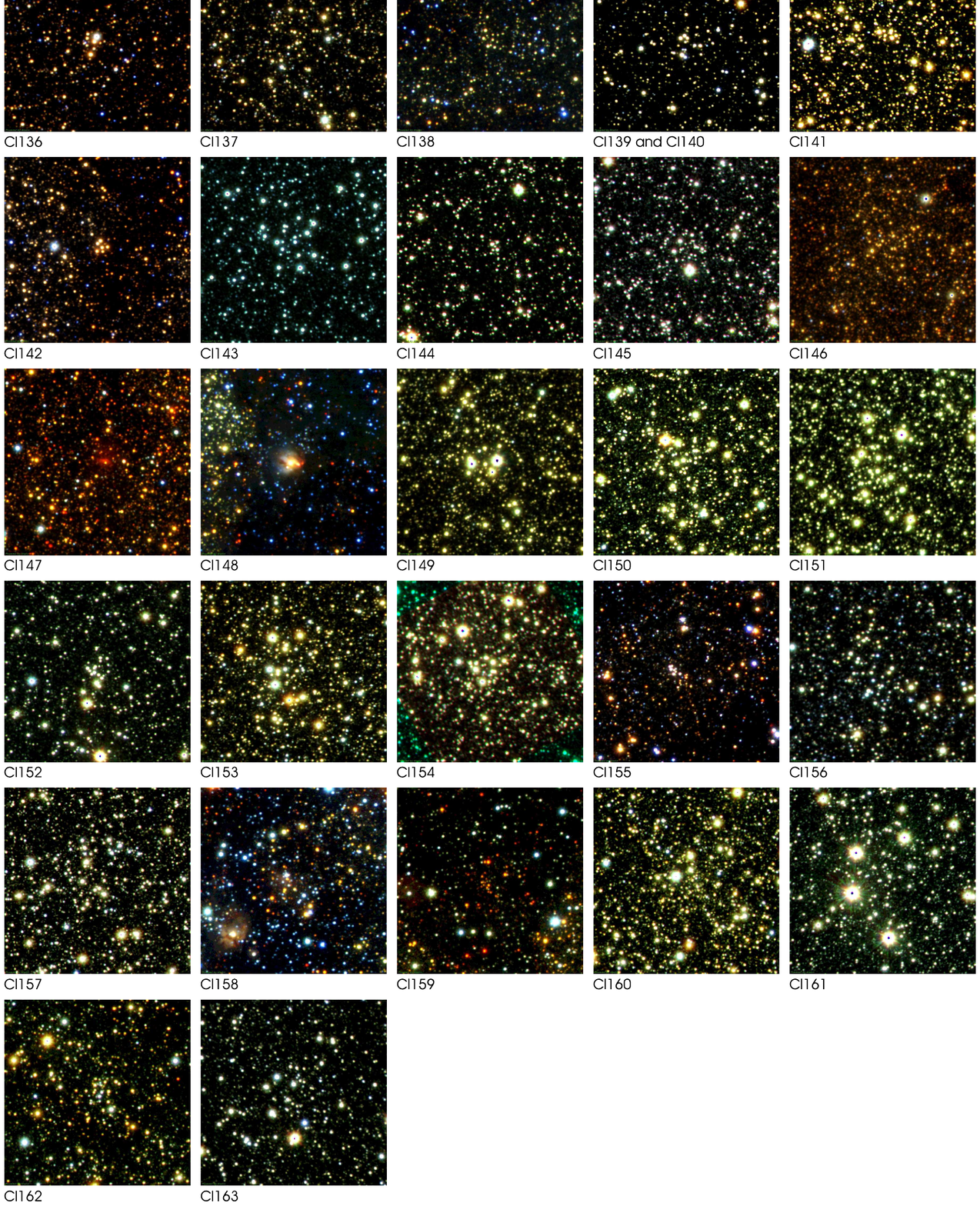}}
\caption{VVV $JHK_{\rm S}$ composite color images of VVV open cluster candidates. The field of view is approx. 2.5$\times$2.5 arcmin, and north is up, east to the right.}
\end{figure}

\newpage
\onecolumn{}  
\begin{figure}[h]
\resizebox{10cm}{!}{\includegraphics{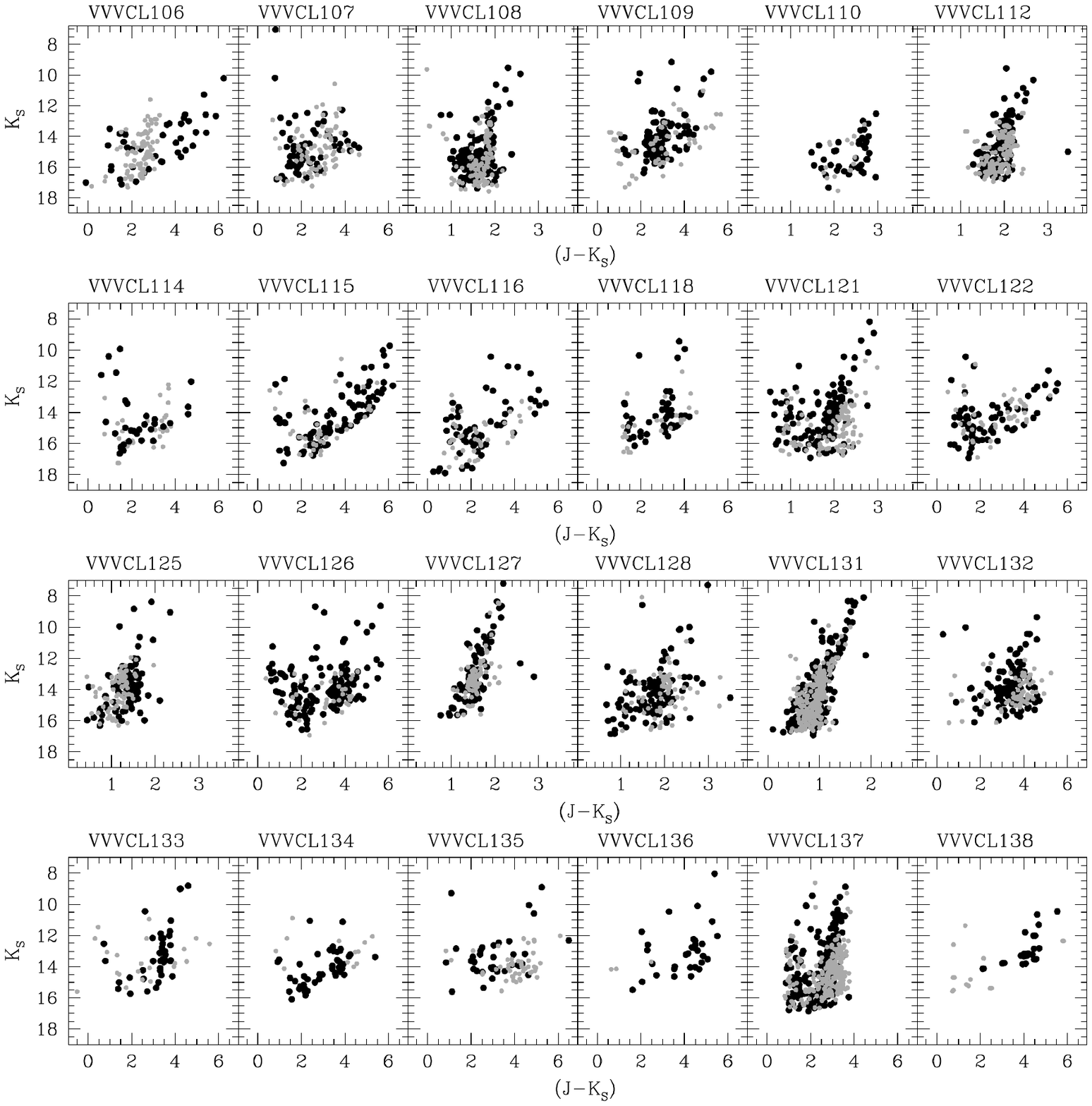}}
\resizebox{10cm}{!}{\includegraphics{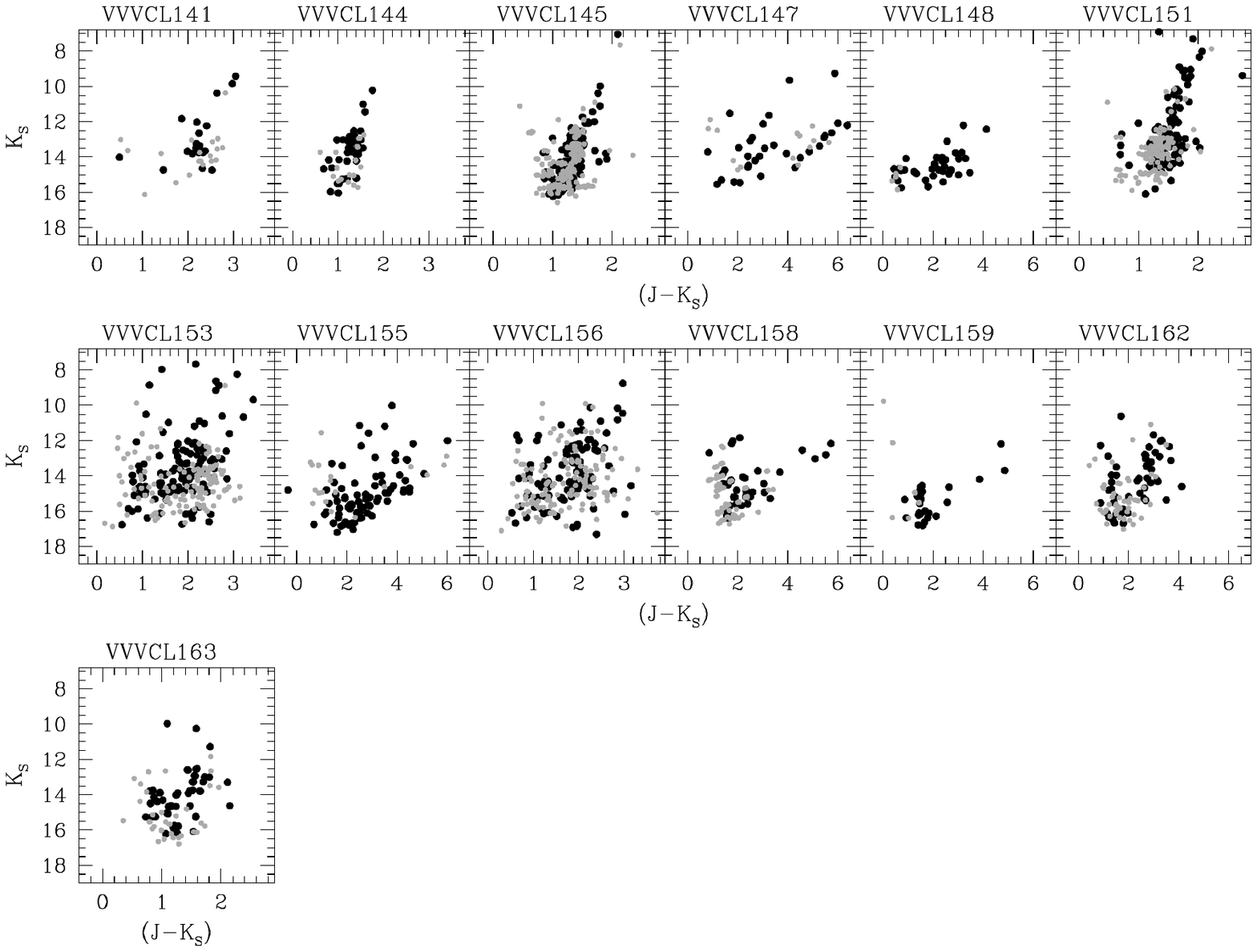}}
\caption{Color-magnitude diagrams of the rest of the clusters listed in Table~1. Gray circles are stars of the comparison field, dark circles are all the stars in the corresponding cluster radius.}
\end{figure}


\end{document}